\documentclass[a4paper,11pt]{article}
\usepackage{graphicx}
\usepackage{amssymb}
\def\wec{wec }
\def\dph{\dot{\varphi}}
\title{New dynamical features of pure  k-essential cosmologies}

\author{Ugo Moschella$^{1,2,3}$ and Mario Novello$^4$ \\ \\
$^1$ Institut des Hautes \'Etudes Scientifiques, \\ 35 Route de Chartres, 91440, Bures-sur-Yvette (France) \\
$^2$ DiSAT, Universit\`a degli Studi dell'Insubria, \\ Via Valleggio 11, 22100 Como (Italy) \\
$^{3}$ INFN, Sez di Milano, Via Celoria 16, 20146, Milano (Italy ) \\
$^{4}$ Centro de Estudos Avan\c{c}ados de Cosmologia/CBPF  \\ Rua Dr. Xavier Sigaud 150, Urca 22290-180 Rio de Janeiro, RJ-Brazil }

\begin{document} 
\maketitle

\abstract{We come back on  the dynamical properties of $k$-essential cosmological models and show how the interesting phenomenological features of those models are related to the existence of boundaries in the phase surface. We focus our attention to the branching curves where the energy density has an extremum and the effective speed of sound diverges. We discuss the behaviour of solutions of a general class of cosmological models exhibiting such  curves and give two possible  interpretations;  the most interesting  possibility regards  the arrow of time that is reversed in trespassing the branching curve.  This study  teaches to us something new about general FLRW cosmologies where the fluids driving the cosmic evolution have equations of state that are multivalued functions of the energy density and other thermodynamical quantities.  }

\section{Introduction}
Models based on a classical scalar field  with  non-canonical kinetic terms, also called  $k$-essential models, were introduced in \cite{damour} and have since then attracted a considerable amount of attention  in cosmology and in modified theories of gravity  \cite{rev2,rev3,rev4}. Because of their innate ability to simulate the behaviour of the cosmological constant,  they can be used to model both the early and the late accelerating phases of the cosmic history and this is why they are widely studied.   Also, they have a remarkable relation with the Born-Infeld theory \cite{novello} and, most of all, with the Chaplygin gas and the associated family of integrable models of fluidodynamics \cite{chap}, relation that enrolls some specific $k$-essential models in the  attempts at unifying the dark  components of the cosmic fluid.  

There are however various types of instabilities that might render some, if not most, of the models in this class pathological. 
One particular instability may described by making use of the  analogy between the scalar field and a fluid, analogy which is valid under the assumption of homogeneity and isotropy, i.e. when  the scalar field depends only on the cosmic time. One may associate to the scalar field an energy density $\rho$ and a pressure $p$ and derive from the Lagrangian a relation between $\rho$, $p $ and the field $\varphi$ that may be called an "equation of state" (eos): 
\begin{equation}
p=p(\rho,\varphi). \label{eoszero}
\end{equation}
The above function is in general ramified: there are  many possible values for the pressure compatible with a given value of the field and of the energy density. Alternatively, the eos  may be introduced in a parametric form:
\begin{equation}
p=p(\varphi,\dph), \ \ \ \ \rho=\rho(\varphi,\dph).
\end{equation}
We may define an effective squared speed of sound as usual, by taking the derivative 
\begin{equation}
c_s^2=\frac {\partial p}{\partial \rho}
\end{equation}
or else  by  the ratio \cite{mu}  
$
c_s^2={\frac {\partial p}{\partial \dph}}/{\frac {\partial \rho}{\partial \dph}}. 
$
The two definitions are in fact equivalent. 

Points where the squared speed of sound diverges define the "critical curves" on the eos surface. Typically they separate  regions where the squared speed of sound is positive from regions where it  is negative; the latter  are usually disregarded \cite{damour,mu} because of a catastrophic perturbative  instability over there.

The state of things  is the same also in the simpler purely kinetic case, where the Lagrangian depends only on the derivatives of the field (actually, only on the time derivative)  and not on the field itself.  
Purely kinetic models have been  be used to describe low energy dynamics of zero-temperature superfluids \cite{1,3,4,5}. Similarly to hydrodynamics these models are subject to formation of caustic singularities \cite{A,B}.  This caustics formation can be avoided by a UV completion. 
 \cite{C,caustics}. This UV completion would also break down around our critical points.  
Here the equation of state is just a curve in the ($\rho,p$)-plane and critical points are the ramification points where the tangent to the curve $p=p(\rho)$ is vertical (see Fig. \ref{figure}). These features were already noticed in the seminal paper \cite{damour} but its explicit focus was only on solutions which describe an expanding universe and therefore the analysis of the dynamical features of the models around critical points was left out.

In what follows we  focus on  the simplest nontrivial  example  of a scalar field minimally coupled to gravity whose dynamics is described by a quartic Lagrangian density 
\begin{equation}
F (\varphi, \partial \varphi  ) =  \lambda(\varphi)  (\partial \varphi)^2 + \mu(\varphi) (\partial \varphi)^4 . \label{theo}
\end{equation}
This model has been introduced in the context of $k$-inflation in \cite{damour} and could also possibly describe the late universe where  dark energy is dominant; our prinicpal aim is contributing to elucidate the role of critical points and surfaces. 

We will at first  in Sect. \ref{sec1} exactly solve  the purely kinetic case $F(\partial \varphi)$ by explicitly constructing the solutions for all possible choices of the relative signs  of  the coupling constants $\lambda$ and $\mu$.  All the interesting phenomena such as the  dynamical generation of either a positive or a negative cosmological constant or the bounce in the cosmological evolution arise as  consequences of the existence of boundaries  in the phase space;  these phenomena are indeed quite model independent and are essentially the same for any choice of a purely kinetic Lagrangian. These facts  are somehow obvious; the minor new points here are the formulae giving the explicit solutions  and a cursory look at a  model mimicking a negative cosmological constant.

In Sect. \ref{sec2}, we focus on the study of branching points  always in the simpler purely kinetic case.  
The presence of a branching point in the physical region of the phase diagram renders the dynamical behaviour much  subtler:  having an explicit solution at hand now is not just an academic  pastime but is crucial to understand what is going on. 

What is puzzling at  the branching point  is the fact that the second time derivative of the field $\ddot \varphi$ diverges while there is no discontinuity in the curvature invariants. We propose two possible ways out of this paradoxical situation.   The more conservative way  requires a discontinuity in $ \dot a(t)$ which, however, is compatible with the absence of discontinuities of the curvature invariants. The evolution of the universe makes the radius shrink with a decreasing  negative velocity $\dot{a}(t)$ till $t=0$ when the universe hits the branching point.

The branching point acts like a wall: the universe undergoes a sort of elastic collision against a wall where the velocity is reverted and becomes positive. Then it enters in a short-lived  phase of decelerated expansion; after a while the expansion stops and the universe  shrinks again until it hits the branching point for a second time. Between the first and the second passage at the critical point the squared speed of sound is negative. However, since the interval is finite the instability may be not catastrophic.

There is however another interpretation   that the field equations seem to dictate: it  points towards the existence of time loops related to the two adjacent branches of the equation of state. In  passing from one branch into the other the sign of time is reversed and the universe bounces and bulges ; once back in the first branch the sign of time is reversed again and the original arrow of time is restored.

We devote the last section of the paper to investigate whether the behaviour of the universe around a critical point we have just described is a peculiarity of  purely kinetic models which are very constrained or else it is a feature that may be found also in more general situations. We show that indeed the latter is the case under quite general condition on the coupling functions  $\lambda(\varphi)$ and $\mu(\varphi)$. 

The new phenomena described in this paper are at the moment mathematical features of pure $k$-essence models. Actually they teach to us something new about general FLRW cosmologies where the fluids driving the cosmic evolution have equations of state that are a multivalued functions of the energy density and other thermodynamical quantities.  Whether there is any observable consequence for the universe we live in is a separate question that requires a better understanding of perturbations around critical points and the presence of other components of the cosmic fluid. We leave this question for further research.

\subsection{A few general facts}
We consider a noncanonical scalar field $\varphi$   described by the Lagrangian density 
\begin{eqnarray}
 && L(\varphi,\partial\varphi) =  F(\varphi, X), \ \ \ \ X = (\partial\varphi)^2 =  g^{\mu\nu}\, \partial_{\mu} \varphi \, \partial_{\nu} \varphi,  
 \label{1}
\end{eqnarray}
minimally coupled to gravity in  a flat FLRW geometry 
\begin{eqnarray}
ds^{2} = dt^{2} - a^{2}(t) \, (dx^{2} + dy^{2} + dz^{2}).
\end{eqnarray}
Homogeneity calls for $\varphi = \varphi(t)$ (so that $X=\dot \varphi^2\geq 0$); 
 the energy-momentum tensor is
\begin{equation} T_{\mu \nu} =  F_X  \partial_\mu \varphi \partial_\nu  \varphi  - \frac 12 F  g_{\mu\nu} 
 \end{equation}
 (in the above formula $F_X = \partial F/\partial X$).
One may think of the field as a perfect fluid whose equation of state is parametrically written in terms of $\varphi$ and its time derivative $\dot\varphi$ as follows:
 \begin{eqnarray}
 && T_{0 0} =    \rho(\varphi,X) 
 = X F_X   - \frac 12 F  
 = \frac 12\left(  \dph \frac{\partial F}{\partial \dph}  -F \right ),  \cr
 && T_{ii} =p(\varphi,X)= \frac 12 F.     \label{rhopi}
\end{eqnarray}
 The field equations  coupling the field to the scale factor are the following: 
 \begin{eqnarray} 
&&  \rho_X  \ddot \varphi +  \sqrt{3\rho} \, p_ X  \dph     +  \frac 12  \frac{\partial \rho }{\partial \varphi } =0 , 
\\
&& 3 \left( \frac{\dot a}{a} \right) ^2= \rho, \label{rhopi2}
\end{eqnarray}
where
$
  \rho_{X}=  X F_{XX}   + \frac 12 F_X ;  
$
 units are such that $8\pi G=1$. 
\vskip 10pt

\begin{figure}[h]\centerline{    \includegraphics[width=12cm]{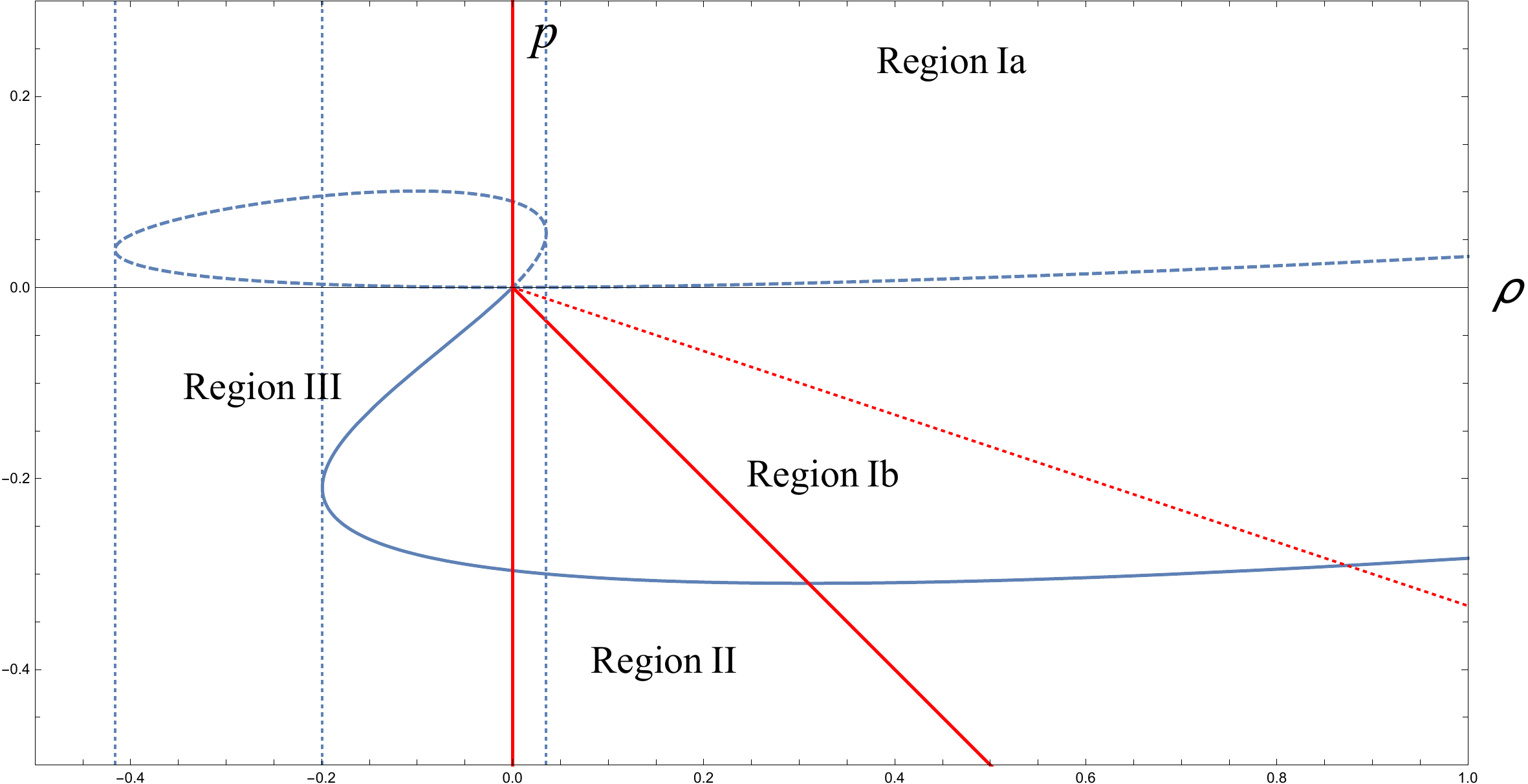}}
 \caption{Plot of the equation of state for the polynomial self-interaction $F(X)=-X-X^2+X^3+X^4$.  }
\label{figure}  \end{figure}

We will at first focus on the purely kinetic case. Here the functions $p=p(X)$ and $\rho = \rho(X)$ define a 
 parametric equation of state that can be represented graphically as a plane curve as we do in Fig. \ref{figure} for a polynomial self-interaction.  The dashed part of the phase curve is unphysical: it is the arc where $X<0$. Also unphysical is the arc of the curve where $\rho$ is negative.

The solid  red lines in Fig. \ref{figure} discriminate three regions : 
\begin{enumerate}
\item{Region I, where the Weak Energy Condition (wec) holds:}
\begin{equation}
\rho \geq 0,\;\;  \ \  \rho +p \geq 0. \label{wecc}
\end{equation}
Region I is subdivided into two subregions: Region Ia where the acceleration $\ddot a(t)$ is  negative ({\em i.e.} $\rho + 3 p > 0$) and Region Ib where it is positive.
\item{Region II, where the Ultra Weak Energy Condition (uwec) holds:}
\begin{equation}
0 \leq\rho \leq - p,  \  \  \  p<0. \label{uwecc}
\end{equation}

\item{ The unphysical region III, where the energy density is negative:}
\begin{equation}
\rho <0 .
\end{equation}
\end{enumerate}
 Region I and Region II are dynamically  disconnected.  Region III is unphysical but dynamically inaccessible since we supposed a flat spatial geometry and the energy density cannot become negative, see Eq. (\ref{rhopi2}).

There  are in general several algebraic branching points where the speed of sound $c_s^2 = \partial p / \partial \rho $ becomes infinite, the energy density has an extremum and the tangent to the phase curve is vertical (see Fig. \ref{figure}). They can be situated either in  the physical regions or in the unphysical one.

 For purely kinetic models  the field equations can be then solved by a quadrature. Since 
\begin{equation}
F_X\,\sqrt{X}  = \pm   \left(\frac {a_0}a\right)^{3}, \ \ \  a_0 \not = 0,
\label{3}
\end{equation}
\begin{equation}
\frac {dX}{dt} = \mp \frac{\sqrt{6} \, X \, F_X \, \sqrt{ X F_X-\frac 12 F}}{2 X F_{XX}+F_{X}} = 
\mp \frac{  ( \rho(X) + p(X))  \, \sqrt{3  \rho(X)}}{\rho'(X)} , \label{xdot} 
\end{equation}
we get 
\begin{equation}
t(X) =\mp \int \frac { \rho_X(X) dX}   {  ( \rho(X) + p(X))  \, \sqrt{3  \rho(X)} } .  
\end{equation}
The only subtlety is about how to glue the solutions around branching points: at a branching point  the density has an extremum,  $\dot X$ diverges and its sign    
may jump. We will clarify this point in the examples.

\section{The quartic purely kinetic model with and without a cosmological constant }
\label{sec1}

\begin{figure}[h] \centerline{ \includegraphics[height=5cm]{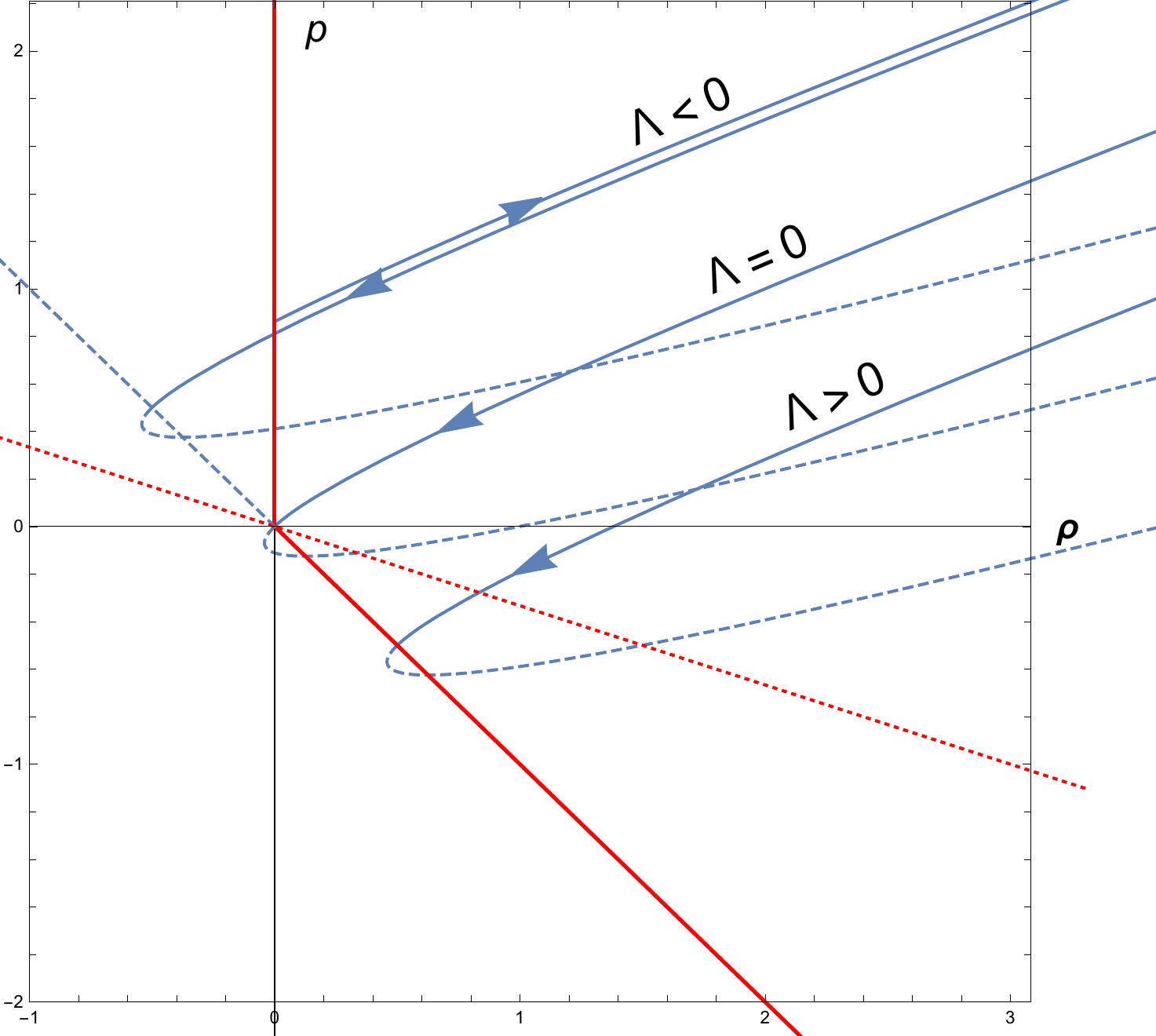} \ \ \  \includegraphics[height=5cm]{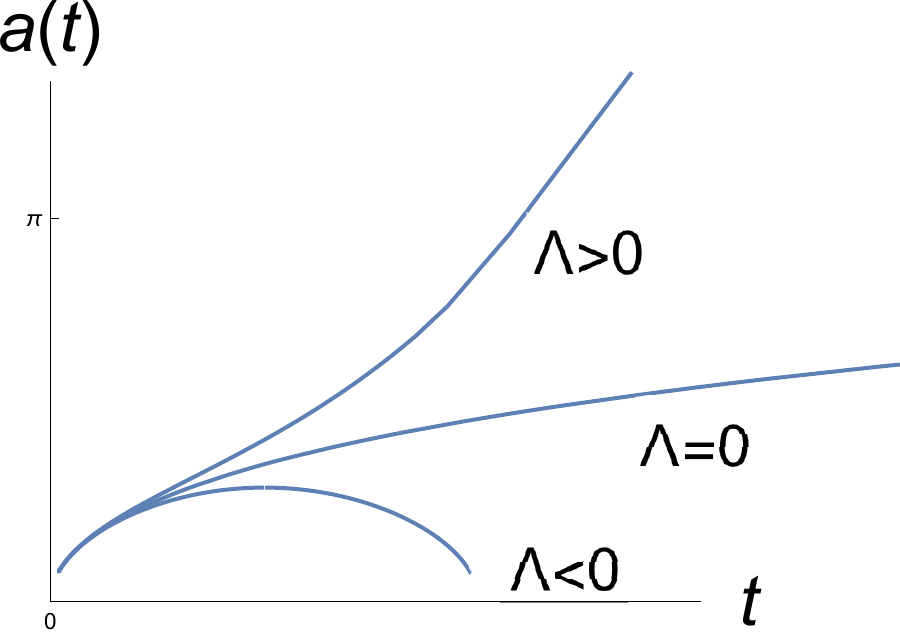}}
  \caption{Plots of the eos and the scale factors for the three possible choices of the cosmological constant. The dashed parts of the phase curves correspond to $X<0$. For $\Lambda\geq 0$  (lower two curves) the wec is respected all the way down to zero,  the  field $X$ takes all the values between infinity and zero and the universe is driven towards the attractor $p = -\rho = -\frac \Lambda 2$. 
For $\Lambda< 0$  (upper curve) not all the positive values of $X$ are attainable. The universe gets to a minimal value of the pressure with zero density and zero velocity; the acceleration is  negative and the universe bounces back and recollapses in a finite cosmic time interval. }
\label{figrp}
  \end{figure}

As a warm-up, let us begin to provide solutions of our prototypical model  (\ref{theo}) in the purely kinetic case by  supposing that  all the coupling constants be  positive;
\begin{equation}
F 
= -\Lambda+   X + X^2 ; \label{theo1} \end{equation}
we  set with no loss of generality $\mu=1$ and $\lambda=1$ and added a cosmological constant term.
%
The physical parts  of the phase curves (where $X>0$) are all contained in region I where the \wec   (\ref{wecc}) is satisfied (see Fig. \ref{figrp}); also,  there are no branching points in the physical region. For  $\Lambda >0$ the solution can be explicitly written in parametric form as follows: 
\begin{eqnarray} 
a(X) &= &{a_0}{(({2 X+1)\sqrt{X}})^{-\frac 13} }, \label{ax2} \\ 
t(X)&=&\frac{\log \left(\frac{2 \Lambda +2 \sqrt{\Lambda } \sqrt{\Lambda +3
   X^2+X}+X}{\left(2 \sqrt{3} \sqrt{\Lambda }+1\right) X}\right)}{
   \sqrt{6\Lambda }}+\frac{4 \log \left(\frac{4 \Lambda +2 \sqrt{4 \Lambda +1}
   \sqrt{\Lambda +3 X^2+X}-4 X-1}{\left(\sqrt{12 \Lambda +3}-2\right) (2
   X+1)}\right)}{\sqrt{24 \Lambda +6}}. \label{cccc}
    \end{eqnarray}
The pressure  becomes negative for $X<\frac 12 (\sqrt {1+4\Lambda} -1) $. After the initial singularity at $t=0$ (i.e. at $X=\infty$) the universe starts decelerating  behaving as as if it were radiation  dominated. The cosmological constant stops the deceleration at $X_\Lambda=\frac{1}{3} \left(\sqrt{3 \Lambda +1}-1\right)$ and it is no surprise that at late times, when $X$ tends to zero, the cosmological constant dominates: $\rho\sim -p \sim \Lambda/2$. The universe evolves between  a radiation dominated phase and an asymptotically de Sitter spacetime. By switching off $\Lambda$ the universe is always decelerating and the model interpolates between radiation at early times and stiff matter $\rho \sim a^{-6}$ at late times (see Fig.  \ref{figrp}) .

 If $ \Lambda = -\Delta < 0 $ the energy density takes negative values in the interval 
\begin{equation}
0< X< {X_\Delta}=\frac{1}{6} \left(\sqrt{12 \Delta +1}-1\right);
\end{equation}
on the other hand  the quantity $\rho + p $ is always positive for $X>0$ . The parametric solution, $a(X)$ does not depend on $\Lambda$  and is the same as in  Eq. (\ref{ax2}); $t(X)$ may be obtained by analytic continuation of Eq. (\ref{cccc}) in the cosmological constant.

It takes a finite interval $t_\Delta$ of cosmic time to get   to $X= {X_\Delta}$ starting from  $X=\infty$; at $X=X_\Delta$ the  density vanishes and the velocity
vanishes as well (this is of course a consequence of the first Friedmann's  equation).  
For a negative cosmological constant the acceleration 
is always negative and therefore the universe bounces back and collapses after $ t_\Delta$ more seconds. The bounce forbids the energy density to become negative but the universe is short-lived.


\vskip 10 pt

 %

 %

Now we get rid of the cosmological  constant  and consider the case $\mu<0$ and  $\lambda>0$.
As before, we may set with no loss of generality $\mu=-1$ and $\lambda=1$
so that 
\begin{equation}
F =  - X + X^2 . \label{lag1}
\end{equation} This  case is discussed in  the seminal paper \cite{damour} and indeed partly explains the interest for $k$-essential models as it catches the most important feature of these models: the cosmic acceleration.
The eos in parametric form is given by \begin{equation} \rho(X)= \frac{1}{2} \left(3 X^2-X\right), \ \ \ \ p(X)= \frac{1}{2} \left(X^2-X\right). \label{eos2}\end{equation}
The  energy density is bounded from below but becomes negative  for $X< \frac 13 $. 
 The  wec is  violated  for   $X< \frac 12 $ as $\rho + p$ becomes negative there.   
There is a branching point at $X_c= 1/6$ but it lies  in the exclusion region. 
There is however a warning \cite{mu}: the squared speed of sound 
\begin{eqnarray}
c_s^2 = \frac{1-2 X}{1-6 X} \label{ss}
\end{eqnarray}
is now negative in the region $\frac 16 <X<\frac 12$. This includes the arc of the phase curve lying in Region II.  

\begin{figure}[h]\centerline{    \includegraphics[height=8cm]{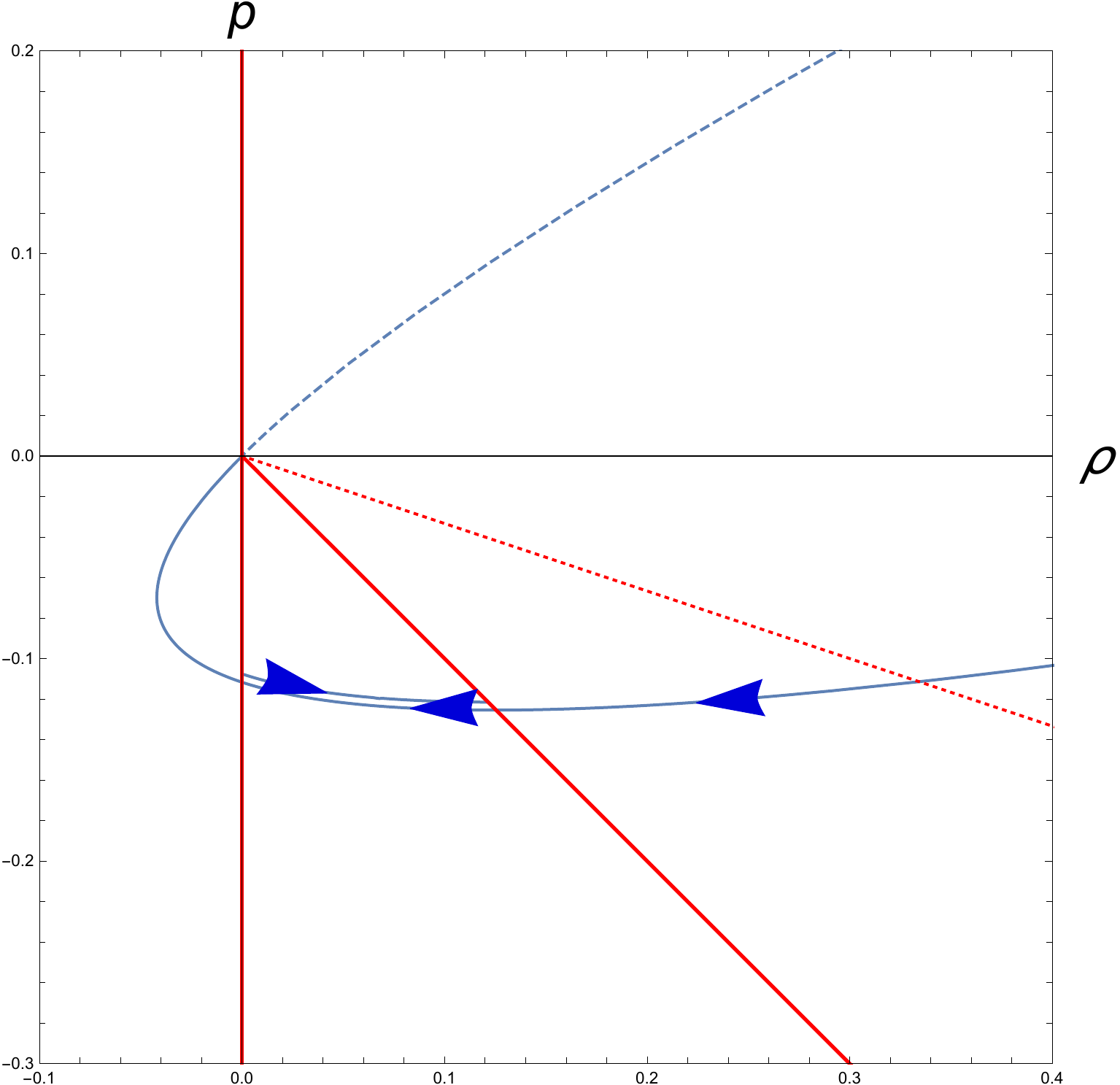}}
  \caption{ Plot of the equation of state (\ref{eos2}).  The branching point is in the exclusion Region III.
 The dashed part of the curve is excluded by the condition   $X>0$. In Region I, after the initial singularity the universe is driven towards the de Sitter attractor  with $p = -\rho = -1/8$  \cite{damour} (compare with  Fig. \ref{figrp}, $\Lambda>0$). On the other hand in Region II  (the uwec region) the universe undergoes a bounce between two asymptotic de Sitter geometries of the same curvature.}
\label{figeos2}
  \end{figure}

A glance at the phase curve in Region I ($X> 1/2$, see Fig. \ref{figeos2}) shows that the dynamical behaviour of this model {\em at the unperturbed level } should be similar  
to the one described above with a positive cosmological constant. However the concavity of the undashed part of the phase curve is opposite to that of 
Fig. \ref{figrp}.
The solution of this model is  written  as follows: 
\begin{eqnarray}a(X) & = &{a_0}{((2X-1)\sqrt{X})^{-\frac 13}},\\
t(X) &= &\frac{8 \sqrt{X (3 X-1)} \tanh ^{-1}\left(\sqrt{\frac{X}{3
   X-1}}\right)+2-6 X}{\sqrt{6X (3 X-1)}}+\sqrt{2}-4 \sqrt{\frac{2}{3}}
   \coth ^{-1}\sqrt{3}.
\end{eqnarray}
 $a(X)$ and $t(X)$   are both defined  and monotonically decreasing in Region I where they are invertible.   They both go  to infinity at $X=1/2$;  values of $X$ smaller than   $1/2$ are    {dynamically inaccessible} (see Fig. \ref{figeos2}).  
The universe  begins with a singularity at $t=0$ ($\dot a = \infty$ and $\ddot a =- \infty$) and immediately enters a radiation dominated decelerating  phase that ends at 
$X=2/3$.
Then the accelerating epoch
starts. 

Both the models (\ref{theo1}) and (\ref{lag1})  interpolate  between radiation at early times 
and the cosmological constant at late times, but the action (\ref{lag1}) contains no  cosmological term; 
at late times (i.e. when $X\sim 1/2$) the quantity $p+\rho $ tends to zero while both the density and the pressure remain  different from zero:
\begin{eqnarray}
\rho(a ) \sim \frac{1}{8} +   \frac{1}{\sqrt{2} a^3} \ \ \ \ \ p(a) \sim \frac 1 8  \  \ \ \ \ \ {\rm for} \    a \to  \infty .  \ \ \ \ 
\end{eqnarray}
 The effective cosmological constant arises dynamically by protection of the wec.  Note also that the speed of sound goes to zero when the universe approaches the attractor.

 \vskip 10 pt

\begin{figure}[h]\center{\includegraphics[height=4cm]{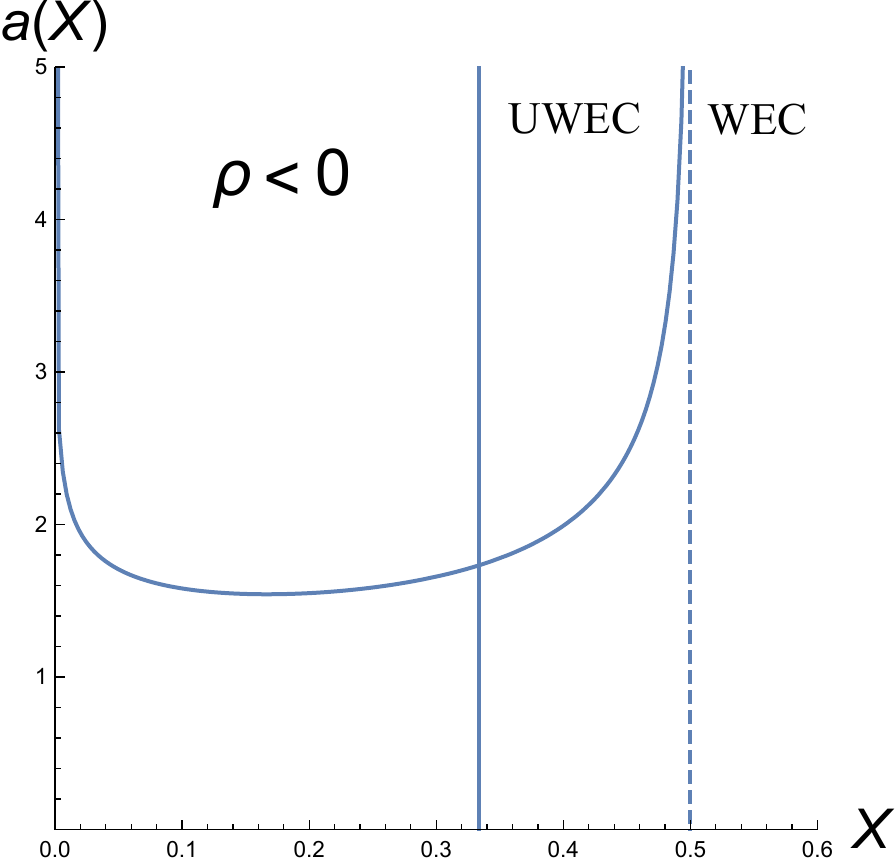} \  \includegraphics[height=4cm]{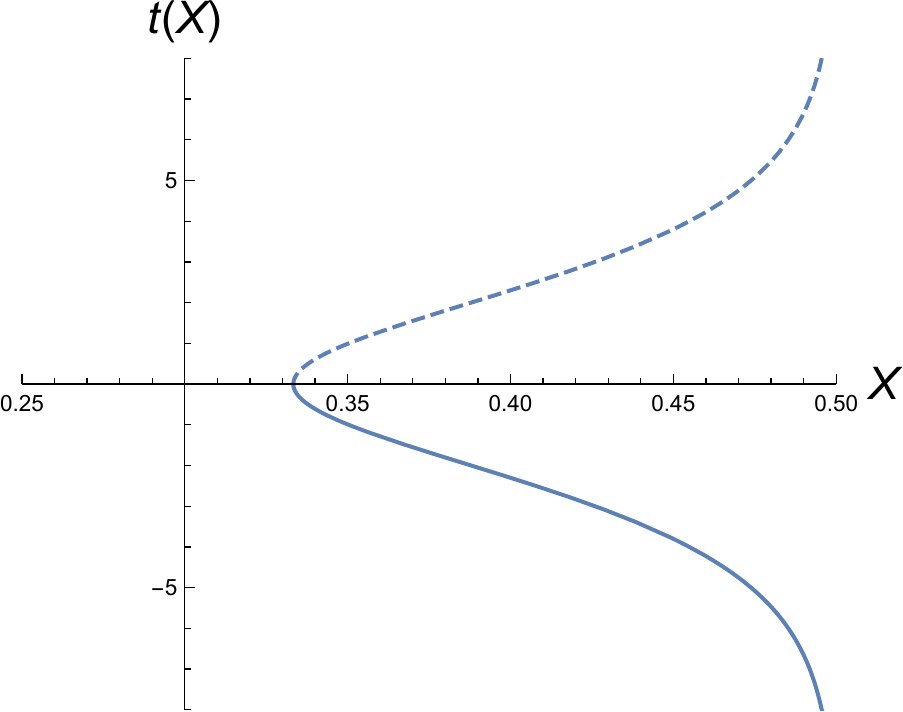} \  \includegraphics[height=4cm]{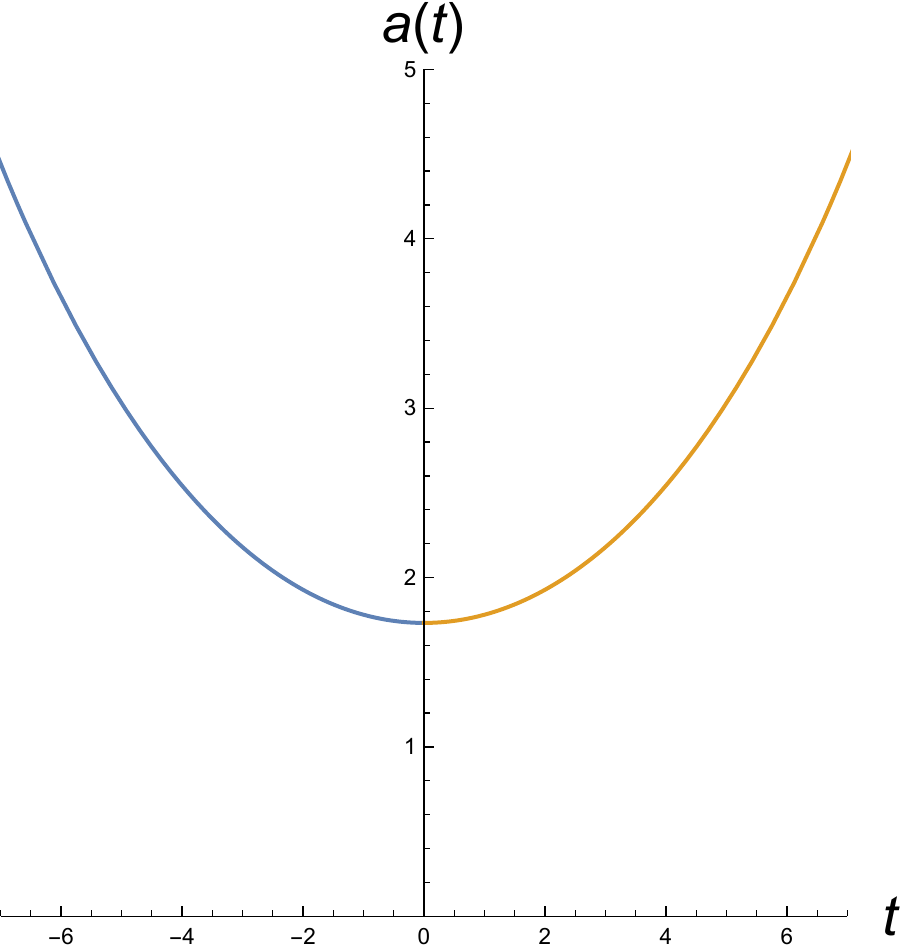}}
\caption{\label{fig21} Plot of the scale factor as function of $X$ and of the time $t$. Here the initial conditions are given in the uwec region $1/3<X< 1/2$. $t(X)$ is a bivalued function of $X$. When it reaches the minimal value $X=1/3$ the universe bounces. }
\end{figure}

The (unperturbed) dynamics becomes a little subtler when the initial condition are given in Region II  (where $\frac 13 <X<\frac 12$).   This case was left out in \cite{damour} because the focus was on expanding universes and most of all because the squared speed of sound here is negative. To construct the parametric solution we should at first choose  $\dot X<0$:
\begin{eqnarray}a(X) & = &{a_0}{((1-2X)\sqrt{X})^{-\frac 13}}  \\
t(X)&=& \frac{6 X-8 \sqrt{X(3 X-1)} \coth ^{-1}\left(\frac{\sqrt{X}}{\sqrt{3
   X-1}}\right)-2}{\sqrt{6X (3 X-1)}}. \label{time}
   \end{eqnarray}
It takes an infinite amount of cosmic time to the universe to get at $X=1/3$ starting from $X= 1/2$; in Eq. (\ref{time}) we have set the arbitrary integration constant so that  $t(1/3)=0$. The function $t(X)$ may be continued as a bivalued function of $X$ by the same formula (\ref{time}) but with opposite sign (the dotted branch in Fig. \ref{fig21}). Inversion provides the function $X(t)$ which is smooth at $t=0$ and therefore provides a smooth scale factor $a(X(t))$. 

The universe starts in a quasi de Sitter contracting  phase at $t\to -\infty$, gets to the minimal value of the scale factor at $t=0$ and then expands asymptotically reaching  the same  expanding de Sitter phase at $t\to \infty$. It is maybe worthwhile to remark that, at variance with the true de Sitter case which would have a spherical geometry, the spatial geometry here is flat. 
This model is however perturbatively unstable because, as we already said,  the squared speed of sound is negative.

 \begin{figure}[h]\centerline{    \includegraphics[width=6cm]{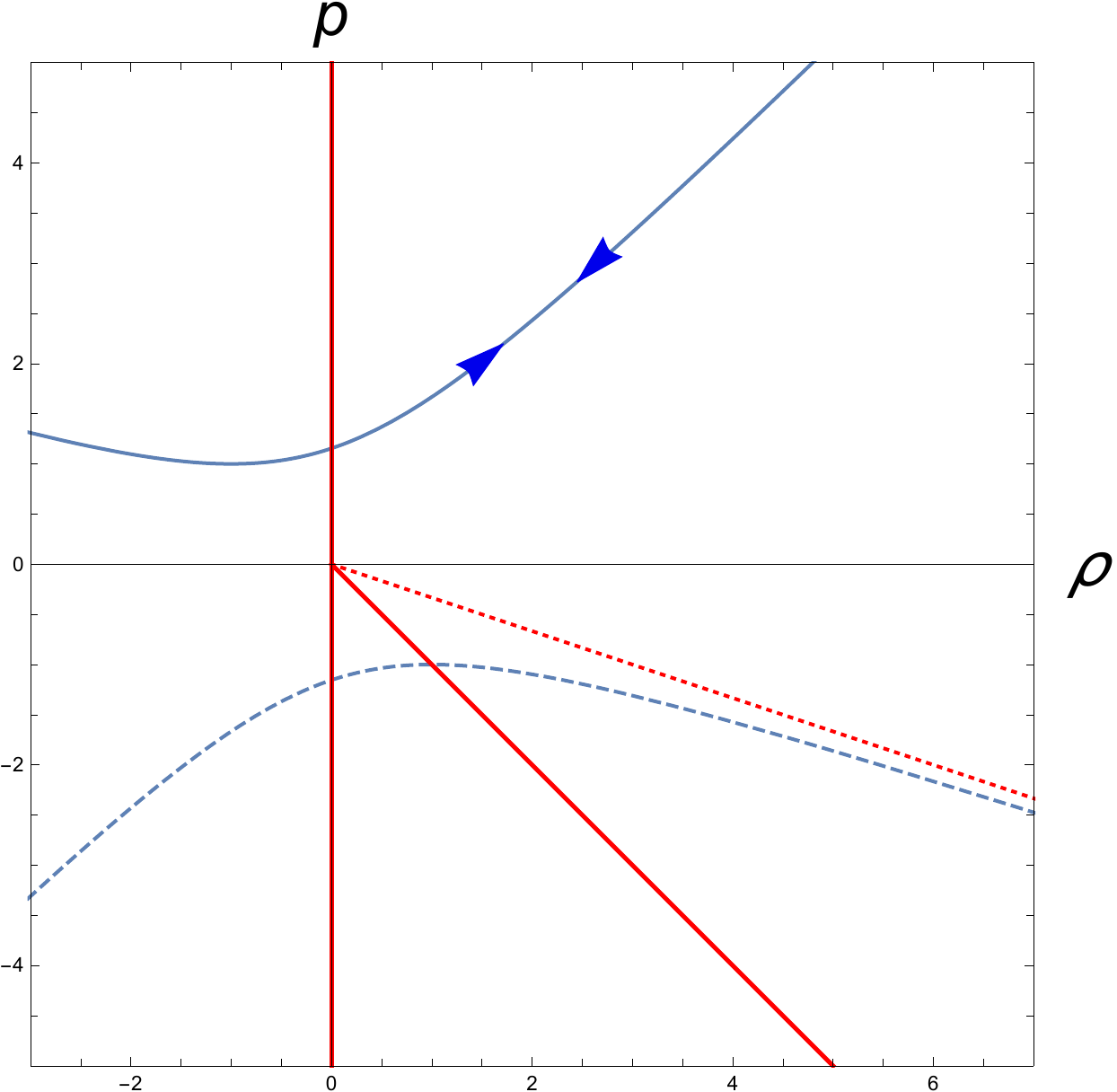}}
  \caption{ Plot of the equation of state (\ref{eos2bb}).  The two branches are disconnected. The branching point is imaginary.
 The dashed branch of the curve is excluded by the condition   $X>0$.}\label{figure9}
  \end{figure}

Is it possible to mimic the dynamical features of a negative cosmological constant in a purely kinetic $k$-essence model?
The answer is yes but it is necessary to go beyond polynomial self-interactions. The simplest model is as follows 
\begin{equation}
F =   X + \frac 1 X  \label{lagneg}
\end{equation}
(see \cite{novello} for a discussion of a similar model of nonlinear electrodynamics). The eos in parametric form is given by 
\begin{equation} 
\rho(X)=\frac{X^2-3}{2 X} ,\ \ \ \ p(X)= \frac{1}{2} \left(X + \frac 1 X \right). \label{eos2bb}
\end{equation}
The  energy density is unbounded from below; it becomes negative  for $X<\sqrt 3 $. 
The  wec is  violated  for   $X<  1 $ as $\rho + p$ becomes negative there.   
 The parametrical solution now reads
\begin{eqnarray} a(X) &= &\frac{ {a_0}\sqrt{X}}{\sqrt[3]{X^2-1}}, \label{ax2bis} \\ 
t(X)&=&\frac 1 {\sqrt{6}} \int_{\sqrt 3}^X \frac{x^2+3}{  \sqrt{x^3-{3}{x}} \left(x^2-1\right)}dx.    \end{eqnarray}

The integral expressing $t(X)$ may be written in terms of hypergeometric and elliptic functions but its explicit expression does not help very much. 
We want just to mention that, as in the negative cosmological constant case discussed in Sect. \ref{sec1},  the lifetime of the universe is finite; here it is given by 
\begin{eqnarray} 
T=\frac{\sqrt[4]{3\pi^2 } \Gamma \left(\frac{1}{4}\right)
   \left(\sqrt{2}+\sqrt[4]{3}
   B_{\frac{1}{3}}\left(\frac{3}{4},\frac{1}{2}\right)\right)}{8 \Gamma
   \left(\frac{7}{4}\right)}.
 \end{eqnarray}
A comment maybe in order here: the finite lifetime arises from the conflict between the tendency of an effective negative cosmological constant to curve the time-like directions and the flatness of the spatial sections.
In this respect, it may be  useful to recall that the purely anti-de Sitter geometry may written as a FLRW geometry only by choosing hyperbolic spatial sections and not as a flat FLRW metric.

\section{Phenomenology of a branching point in the purely kinetic case and the arrow of time}
\label{sec2}
\subsection{The energy density has a maximum value}
Here we start describing to the new features that are investigated in this paper i.e. the dynamical  behaviour of a model universe around branching points.  These features are at first described in a purely kinetic model 
whose solution can be explicitly displayed:
%
\begin{eqnarray}
 F =  X - X^2, \ \ \ \ \rho= \frac{1}{2} \left(X-3 X^2\right), \ \ \ \ p= \frac{1}{2} \left(X-X^2\right), \ \ \ \ c_s^2 = \frac{1-2 X}{1-6 X} \label{eos3}
\end{eqnarray}
(we set 
$\mu=1>0$ and $\lambda=-1<0$).

In this model the energy density has a maximum value, becomes negative  and  is not bounded from below. 
In flat spacetime the model would be considered  pathological but the curvature may improve its status. 

\begin{figure}[h]\centerline{ 
  \includegraphics[height=6cm]{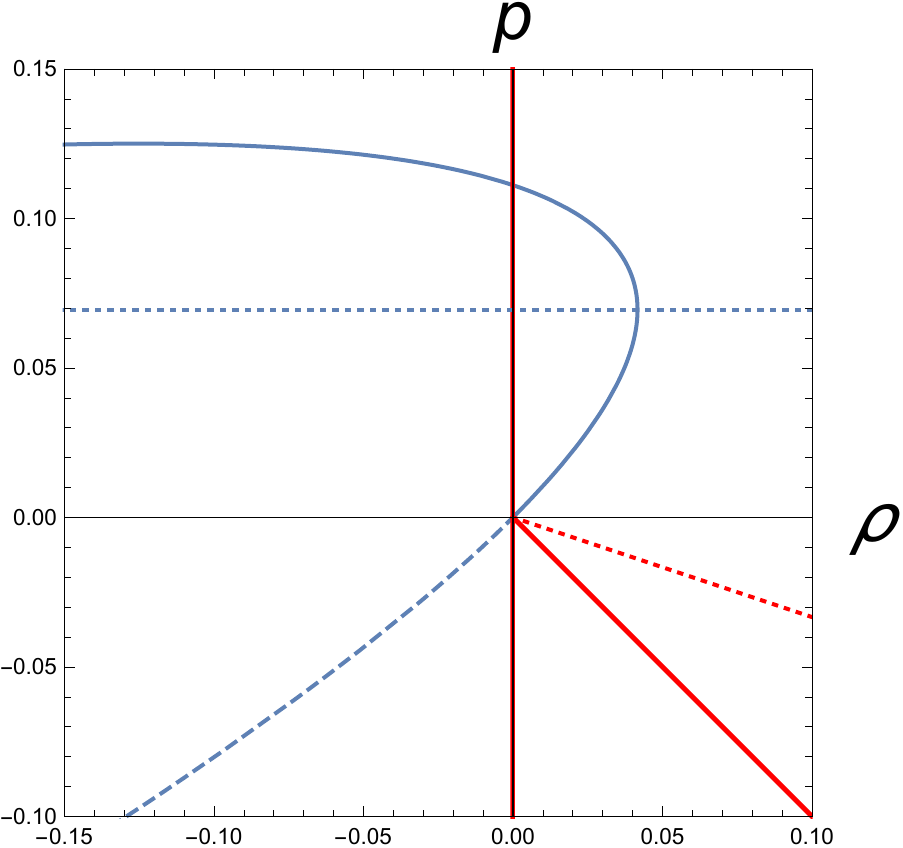}}
  \caption{The branching point is in Region Ia.  At the branching point the energy density reaches its maximum but $p_X\not = 0$;  the  tangent to the phase curve is vertical.  }
\label{figeos3}
  \end{figure}

The squared speed of sound is the same as in Eq. (\ref{ss}), being invariant w.r.t. a global change of sign of the Lagrangian. As before it is  negative when  $\frac 16 <X<\frac 12$. This now  includes the  arc of the upper branch of the phase curve contained in the first quadrant of Fig. \ref{figeos3}, where the energy density is positive.  

The branching point  is at  $X_c= 1 /6$  where the  speed of sound diverges and the energy density reaches its maximum (see Fig. \ref{figeos3}). 
The condition of criticality $c_s^2=\infty$ is thus equivalently written as follows:
\begin{equation}
\frac {\partial \rho}{\partial X}=0, \ \ \ \ \ \ \frac {\partial p}{\partial X}\not =0.
\end{equation}
The case where the above  derivatives are both zero (with the coupling constants that may also depend on the field $\varphi$, see below) plays a central role to explore the possibility of a transition to a phantom regime in the model (\ref{theo})  in \cite{Vikman}; in our case however $p_X$ does not vanish at the critical point.

The critical point is a ramification point for the equation of state; the upper and lower branches are respectively given by 
\begin{equation}
p= \frac{1}{18} \left(6 \rho+1 \pm \sqrt{1-24 \rho }\right).
\label{rami}
\end{equation}
The presence of a branching point in the physical region renders the dynamical behaviour much  subtler and  care is required to understand it. 
Here, having an explicit solution at hand is not just an academic play but a crucial help in understanding the dynamical behaviour of the model.
The equations are 
\begin{eqnarray}
a(X)&=& a_0 \left({{\sqrt{X}(1-2 X)}} \right)^{-1/3}\label{ax3},  \\
  \dot X &= &\pm\frac{\sqrt{6} X (1-2 X) \sqrt{X-3 X^2}}{1-6 X} \label{eosi3}
\end{eqnarray}
and the  task is to describe  the scale factor $a(t)$ as a function of the cosmic time. 
\subsubsection*{Lower branch of the state equation}

\begin{figure}[h]
\center{
 \includegraphics[height=4.5cm]{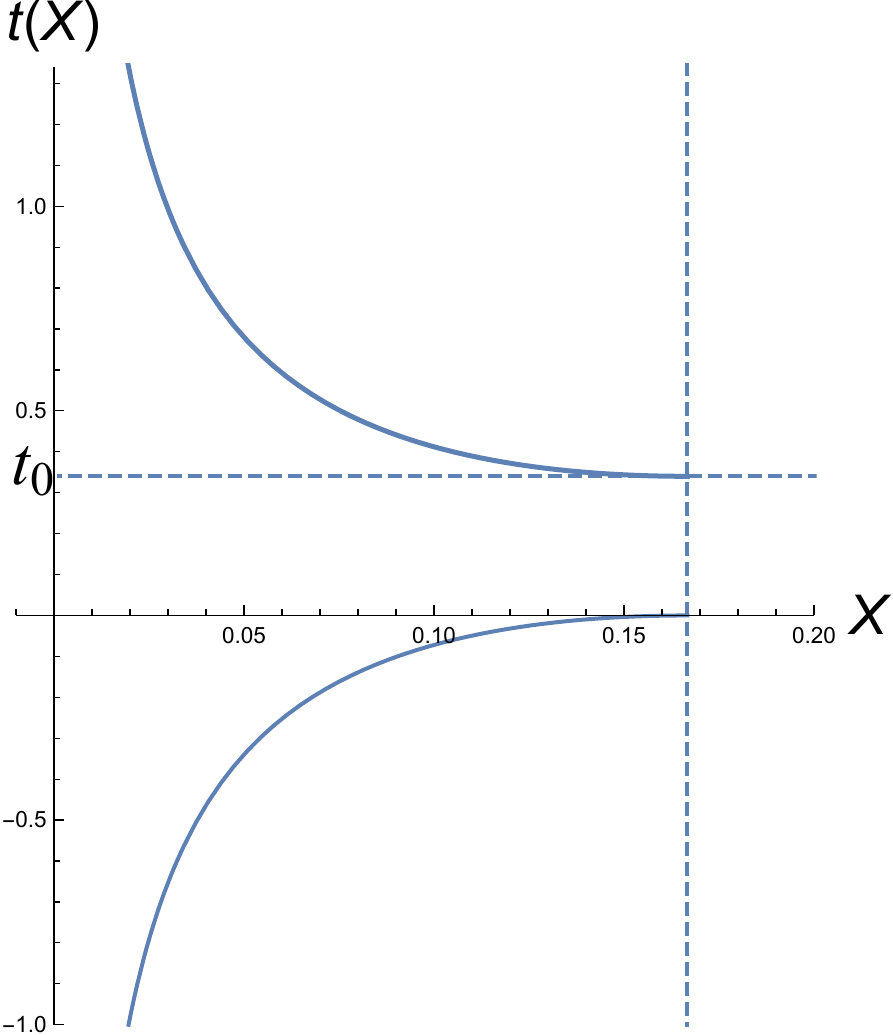}    \ \ \ \ \ \ \ \includegraphics[height=4.5cm]{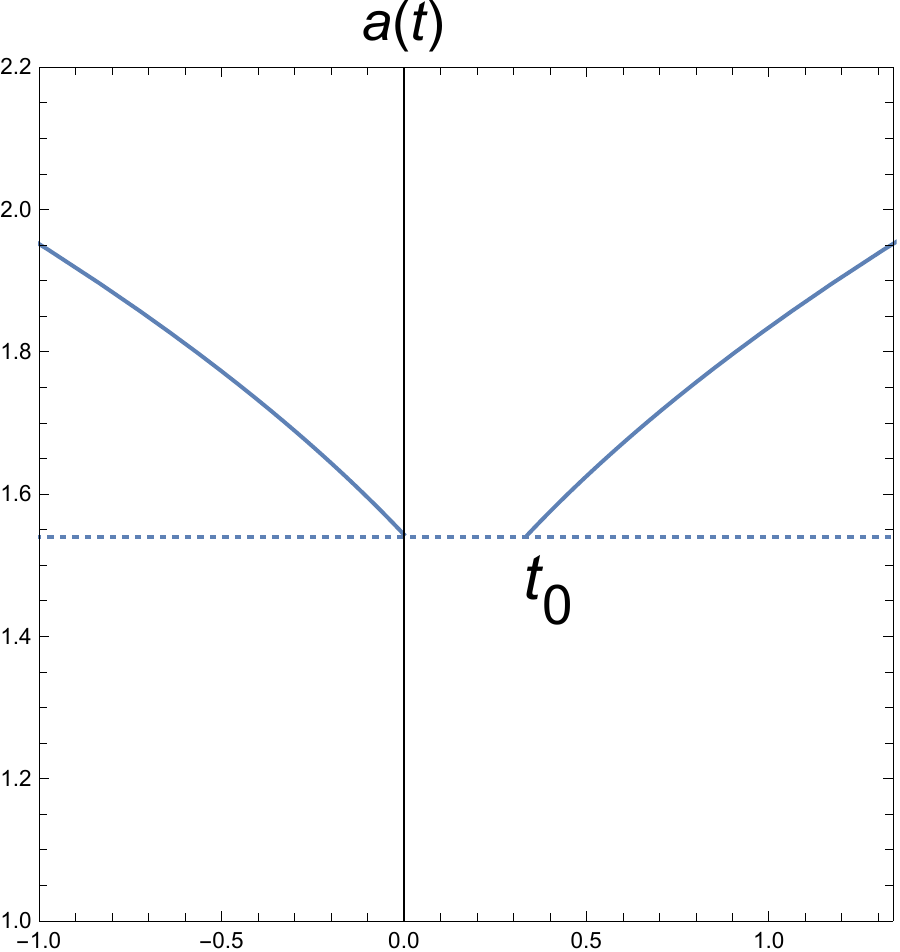}  } 
\caption{\label{Fighe} Approaching the branching point and stepping away from it from the left $X<X_c$.}
\end{figure}

Let us start at $X\sim 0$ (which means $a\sim \infty$). 
$X$ cannot become negative  and therefore at this stage we have to choose $\dot X >0$ ({\em i.e.} the plus sign in Eq. (\ref{eosi3})).
With  this choice $\dot a$   is a negative decreasing function of $X$ in the interval $0< X<X_c$;  the acceleration  is also negative
since these points are in region Ia:
\begin{eqnarray}
\dot a(X) &=& -\frac{\sqrt{X-3 X^2}}{\sqrt{6} \sqrt[3]{(1-2 X) \sqrt{X}}}= - \sqrt{\frac{\rho(X)}3 } a(X) \label{acc}
\\
\ddot a(X)&=&- \frac{3 X (2-3 X)}{2 \sqrt[3]{(1-2 X) \sqrt{X}}} = -\frac { a(X) }6  (\rho + 3 p)  < 0  \label{Rayc}
\end{eqnarray} 
(here we set $a_0 = 1$). The universe gets to the branching point with finite, strictly negative velocity and acceleration: 
\begin{eqnarray}
a(X_c)=\frac{\sqrt{3}}{\sqrt[6]{2}},\  \ \ \  \dot a(X_c)=-\frac{1}{ 2^{5/3} \sqrt{3}} , \ \ \ \  \ddot a(X_c)= -\frac{1}{8 \times {2}^{\frac 1 6} \sqrt{3}}. \label{values}
\end{eqnarray}
All the curvature invariants are finite and nothing special seems to happen there. Integrating  Eq. (\ref{eosi3}) up to the branching  point gives:
\begin{eqnarray}
 t(X) =\frac{6 X-2-8 \sqrt{3 X^2-X} \tanh ^{-1}\left(\frac{\sqrt{X}}{\sqrt{3
   X-1}}\right)}{\sqrt{6 X-18 X^2}}+\frac{2}{3} \sqrt{\frac{2}{3}} \pi
   +\sqrt{2}, \ \ \ 0<X<1/6,  \cr \label{uhu}
   \end{eqnarray}
where  we set $t (X_c)= 0$. The tangent to the curve $t(X)$ at $X_c$ is horizontal (see Fig. \ref{Fighe}).  We may also give initial conditions at the branching point by reverting  the velocity  in Eq. (\ref{values}) i.e. 
\begin{equation}
a(X_c)=\frac{\sqrt{3}}{\sqrt[6]{2}},\  \ \ \  \dot a(X_c)=\frac{1}{2^{5/3} \sqrt{3}};  \label{values2}
\end{equation}
in this case we should choose the negative sign  in Eq. (\ref{eosi3}) and the solution is given by 
$
 t_1(X) =t_0 -t(X), \ 0<X<1/6 ,
$
where a suitable value of the constant $t_0$ has still to be chosen. These functions and the corresponding scale factors $a(t)$ are plotted in Fig. \ref{Fighe}. 
\begin{figure}[h]
\center{\includegraphics[height=4.5cm]{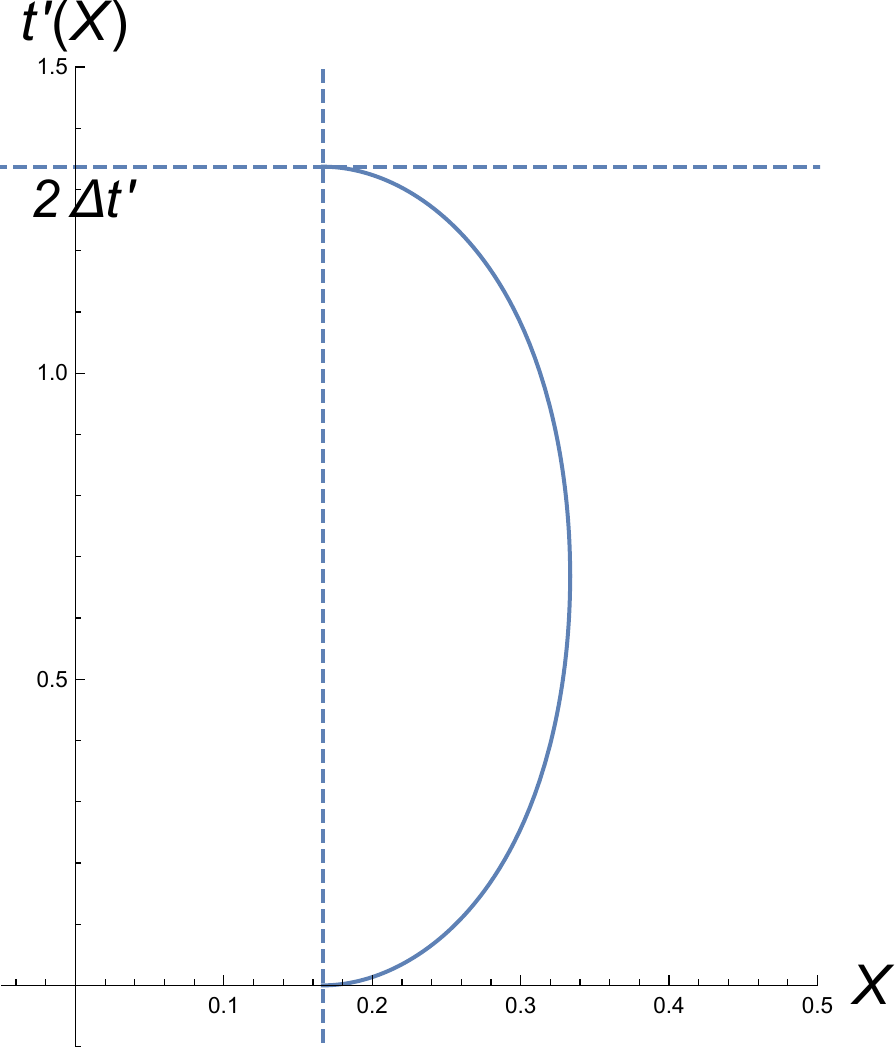}    \ \ \ \ \ \ \ \includegraphics[height=4.5cm]{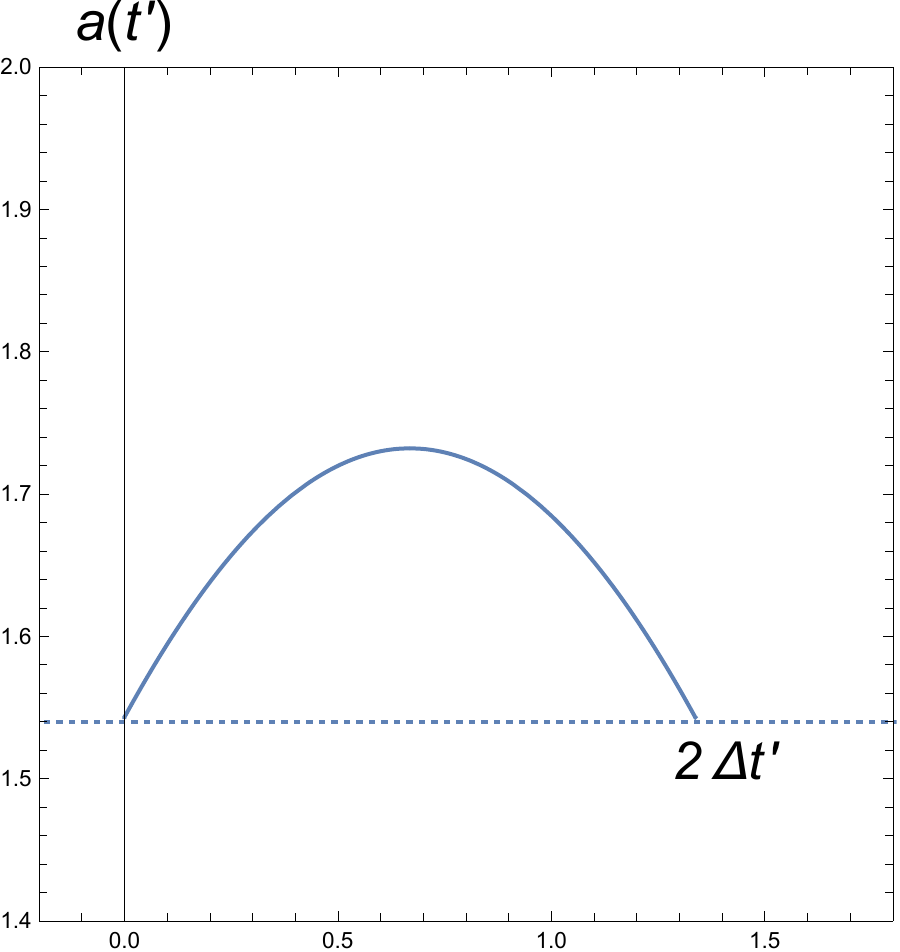}  } 
\caption{\label{Fighe2} Stepping away from the branching point, bouncing  at the boundary of the allowed region and approaching the branching point from the right  $X>X_c$.}
\end{figure}

\vskip 5pt
The question that is not completely trivial remains unanswered yet: once at the branching point, what happens next?
To give a possible solution to this problem we need to solve the model for the (unstable, since $c^2_s<0$ on it) upper branch of the equation of state.

\subsubsection*{Upper branch of the state equation}
We may  set the same initial conditions (\ref{values2}) at the branching point but  consider instead  a positive $\dot X$;  this amounts to the negative sign  in Eq. (\ref{eosi3}) because we are exploring the region  at the  right of the critical point $X> 1/6$.  
Integration gives 
\begin{equation}
t'(X)=\frac{8 \sqrt{X-3 X^2}  \tan
   ^{-1}\left(\frac{\sqrt{X}}{\sqrt{1-3 X}}\right)+2-6 X}{ \sqrt{6X-18
   X^2}}-\frac{2}{3} \sqrt{\frac{2}{3}} \pi -\sqrt{2}
 \end{equation}
The universe expands  from $X_c=1/6$ to $X= 1/3$ in a finite time:
\begin{equation}
\Delta t '= t'(1/3)=\frac{1}{9} \sqrt{2} \left(4 \sqrt{3} \pi -9\right).
\end{equation}
At $X=1/3$  the scale function has a maximum, the velocity vanishes and the acceleration is strictly negative. The solution is smoothly continued by 
$
t'_1(X)= 2\Delta t' -t(X).
$
The universe start recollapsing and get back to the branching point in another $\Delta t'$ seconds; at $X_c$ the scale factor and its derivatives have the values given in Eq. (\ref{values}) (see Fig. \ref{Fighe2}).

\subsubsection*{Matching of the lower and upper solutions}

\begin{figure}[h]
\center{\includegraphics[width=6cm]{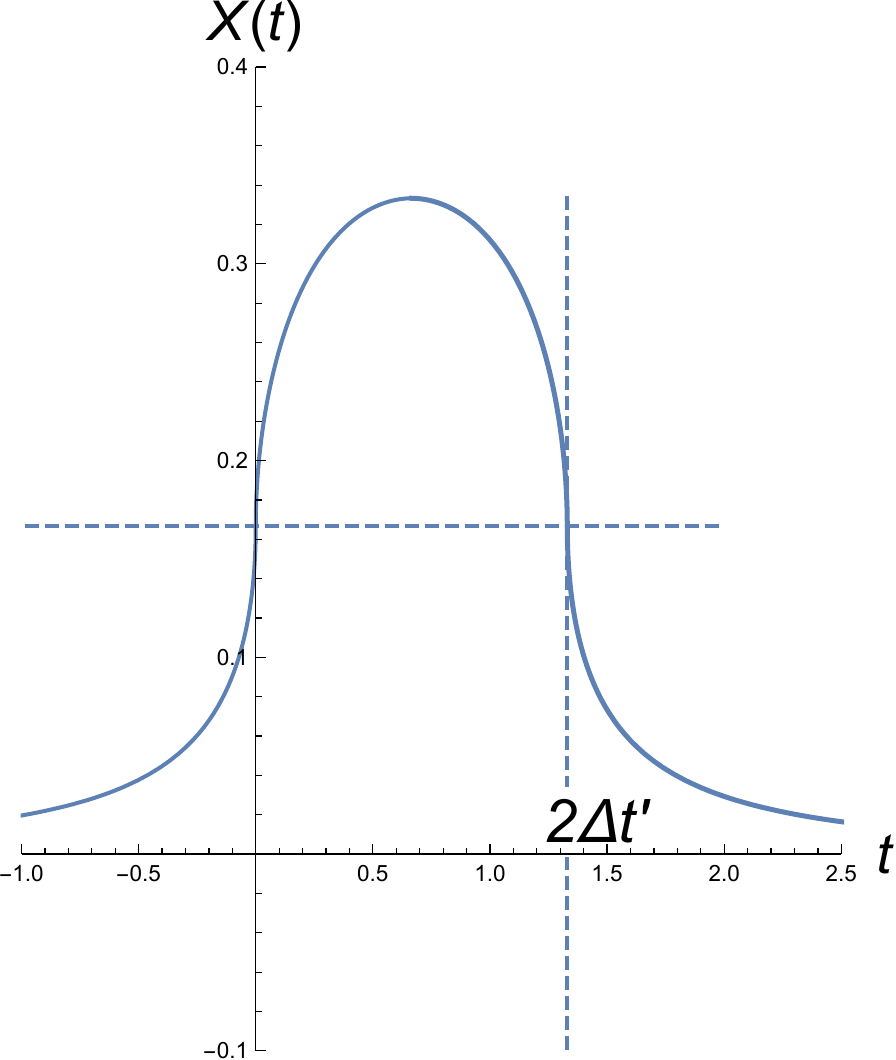} \ \ \ \ \  \includegraphics[width=6cm]{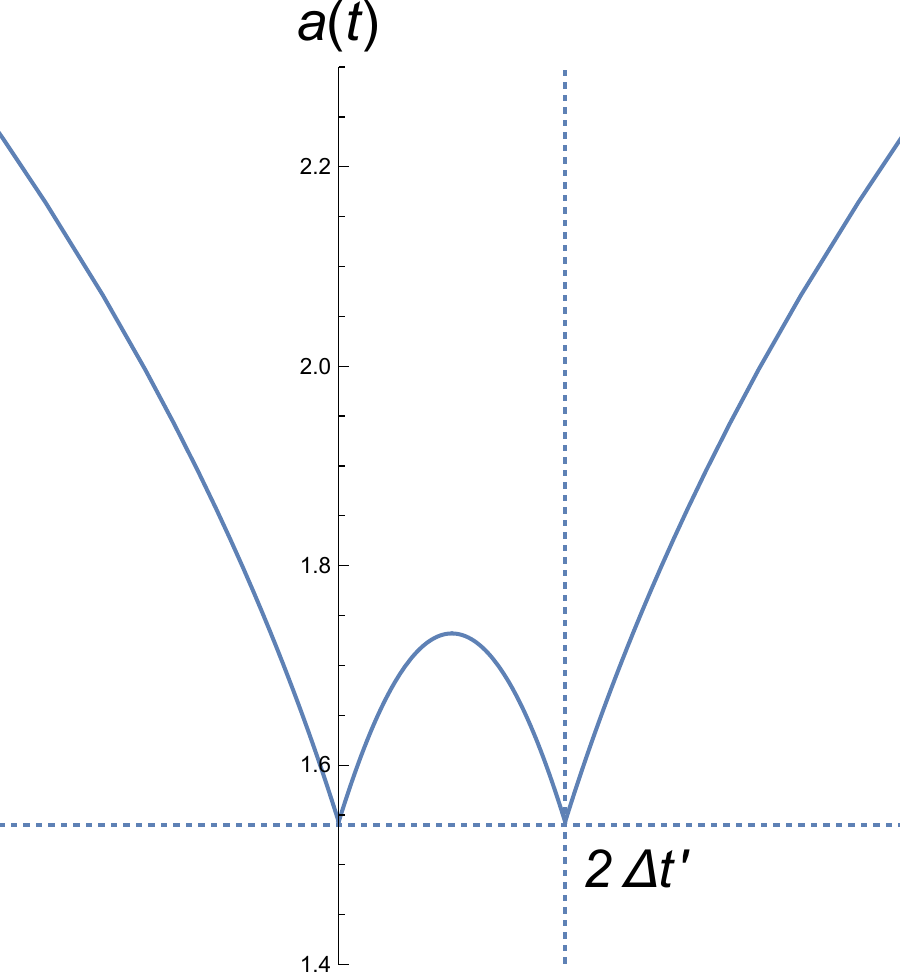}       }
\caption{\label{fig46} Plot of the field $X(t)$ and the scale factor $a(t)$  as  functions of the cosmic time. The are both  irregular at $t=0$ and $t= 2\Delta t'$.} \label{fighe3}
\end{figure}

There is a conservative way to match the upper and lower solutions described above: they are glued together simply by choosing\footnote{We set aside the  more artificial and less interesting possibility to chose $t_0=0$.}
$t_0 = 2 \Delta t'$ This choice amounts to thinking of the branching point as behaving  like a wall.
When the field  gets at $X_c$ the density reaches a maximum and the universe cannot shrink anymore. Therefore, like in the elastic collision of a ball against a wall, the sign of $\dot a$ is reverted; on the other hand  the sign of $\dot X$ does not change while  traversing the critical point. The expansion following the collision  keeps decelerating as the acceleration in this model is aways negative. 
\vskip 5 pt
The scale factor $a(t)$ is continuous but not differentiable at the the branching point (see Fig. \ref{fighe3}). The left and right derivatives have the same modulus but opposite signs.
As regards the acceleration, the left and right derivatives of the velocity w.r.t. $X$ vanish at the branching point but, when multiplied by the diverging function $\dot X$, they have the same limit 
and Eq. (\ref{Rayc}) is still valid.
Also the curvature invariants, which depend only on $\dot a^2$, have no discontinuity at the branching point.

\subsubsection*{History of the universe}
\begin{enumerate}
\item The Universe starts at $t=-\infty$ with infinite radius and zero velocity. It is a flat Minkowski spacetime.
\item The evolution of the universe makes the radius shrink with a negative and decreasing velocity $\dot{a}$ up to $t=0$ when the universe hits the branching point.
\item {\em The (unstable) bulged bounce}. The branching point acts like a wall and the universe undergoes an elastic collision where the velocity is reverted and becomes positive. The universe  enters in a phase of decelerated expansion. At $t = \Delta t'$ the expansion stops and the universe starts again to shrink up to  $t = 2\Delta t'$ where it hits the branching point  for a second time. In this phase the squared speed of sound is negative. However it lasts a finite interval of time that may be very short; therefore the instability may be not catastrophic.
\item At  the branching point the velocity is reverted again. The universe enters in a phase of everlasting decelerated expansion that will drive it back to Minkowski space at $t=\infty$.
\item There is no singularity at $t=\infty$.
\end{enumerate}

\subsubsection*{Back to the future. Matching of the lower and upper solutions reloaded }

There is however a drawback in the previous construction due to the stretch we made in gluing the solutions as we did.
Indeed, the non-differentiability of the function $a(t)$ w.r.t. to the cosmic time arises because of the conservative choice we made in insisting that the time variable unfolds regularly and always in one direction.
This choice should however imply the presence of distributional  delta contributions at the branching point, something that  we avoided  by joining the left and the right limits of the scale function and its derivatives at the branching point.  This would be a junction condition different from the Darmois and Lichnerowicz's junction conditions; the comforting aspect of this procedure is that in doing so no discontinuity arises in the curvature invariants.

There is however an alternative viewpoint that  automatically emerges by taking  the solution of the field equations at its face value  without any stretch: the solution (\ref{uhu}) is actually well defined in the whole interval $0<X<1/3$ and no effort of imagination  is needed in taking it as such and joining it with the reflected solution $t_1(X) = -2\Delta t' -t(X)$, also valid in the full region $0<X<1/3$.
\begin{figure}[h]
\center{\includegraphics[width=6cm]{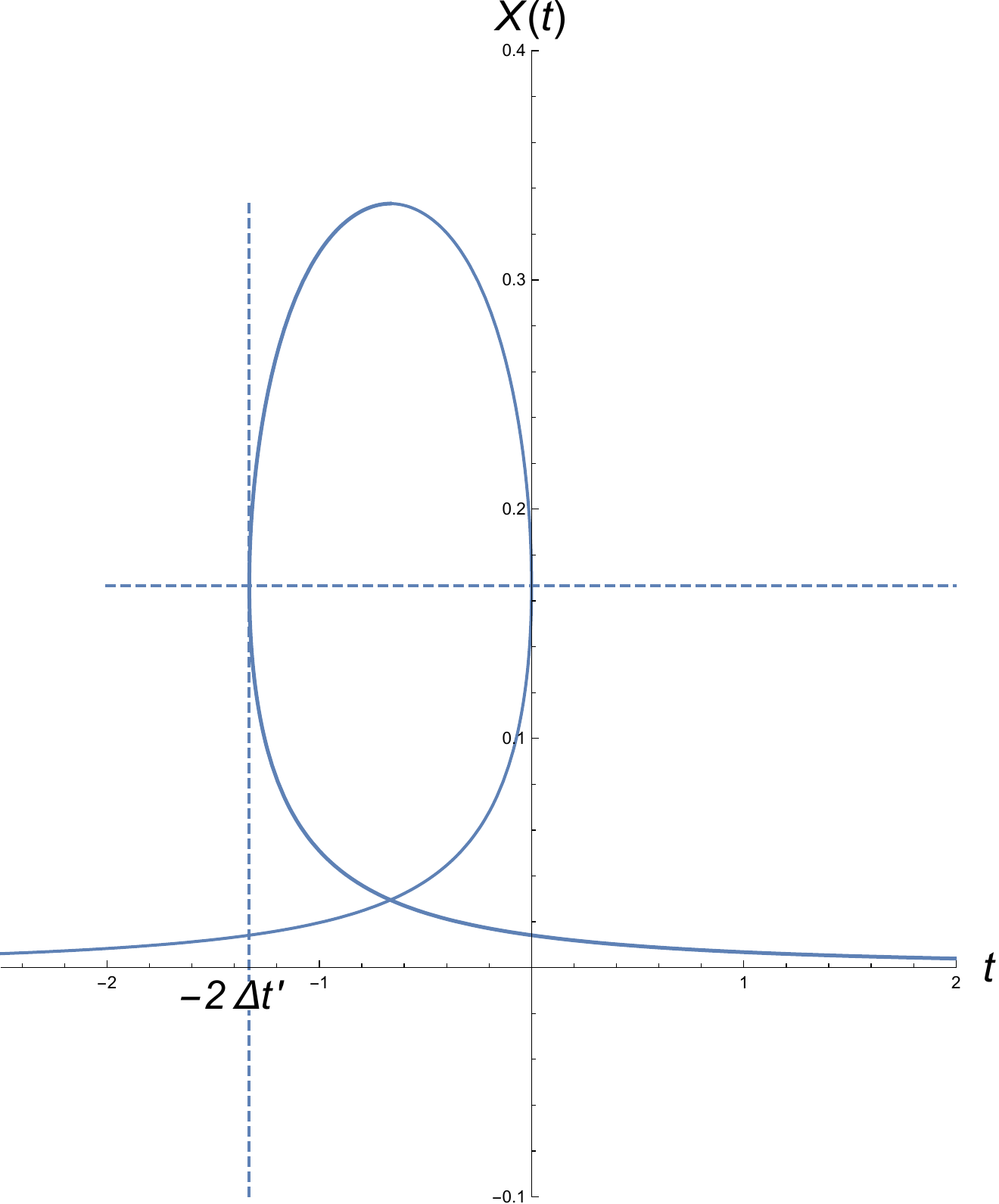} \ \ \ \ \  \includegraphics[width=6cm]{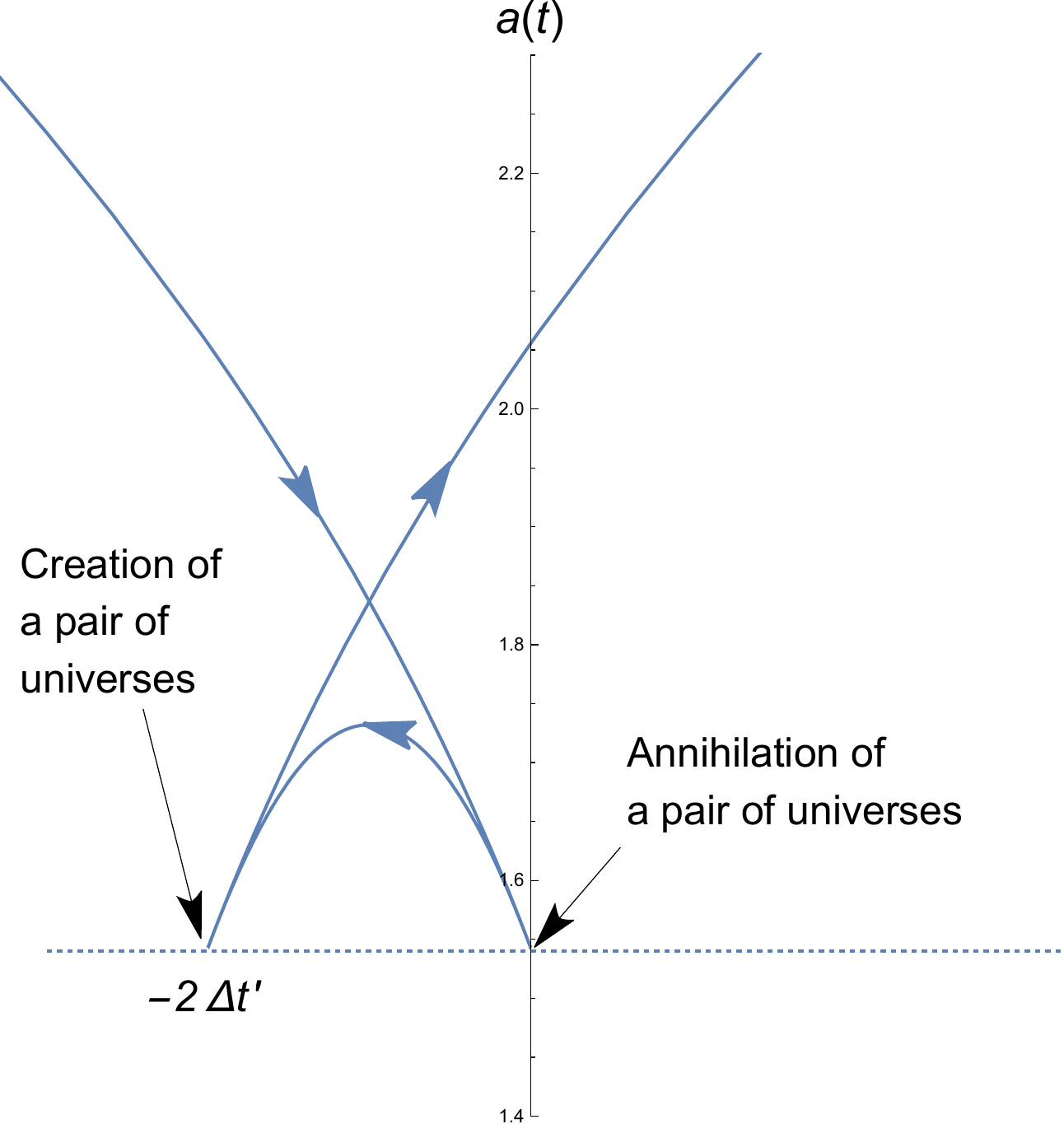}       }
\caption{\label{fig46bis} Plot of the field $X(t)$ and the scale factor $a(t)$  as  functions of the cosmic time. They are multivalued but regular at the branching times  $t=0$ and $t=- 2\Delta t'$.} \label{fighe4}
\end{figure}

The solution $X(t)$ and $a(t)$ are plotted in {\em trompe l'\oe il } w.r.t. the time variable in Fig.  \ref{fighe4}.  What is the meaning of such diagrams? They represents the clever solution that the universe gives to apparently unsolvable problem it has to face when arriving at the branching point: how could it go from 
$a(X_c)=\sqrt{3}/{\sqrt[6]{2}}$ to $a(1/3)= \sqrt{3}$ with a negative velocity and from $\dot a(X_c=1/6)=-{\sqrt{3}}/{ 2^{5/3}}$  to $\dot a(1/3)=0$ with a negative acceleration? Running backward in time!

The ramification point in the equation of state is encountered twice during the time evolution. When the universe gets at the ramification point the time starts flowing backward till the universe gets again to the ramification point, when the usual forward orientation of time is recovered. The phenomenology is essentially the same as in the previous conservative description but the inversion of the velocity is caused by an inversion of the sense of the flow of time, inversion that is short-lived.

Borrowing relativistic quantum mechanics ideas of Dirac, Stuckelberg and Feynman, it is tempting to say that a universe pair is annihilated when the ramification point is first encountered; a pair of universes is created when the ramification point is encountered a second time.

Here the time runs backward for a finite interval that may be very short; the issue of the instability due to the negative squared velocity of sound needs therefore to be reconsidered. We leave this investigation for future work.

\subsection{Energy has a minimum}
\begin{figure}[h]\centerline{ \includegraphics[width=6cm]{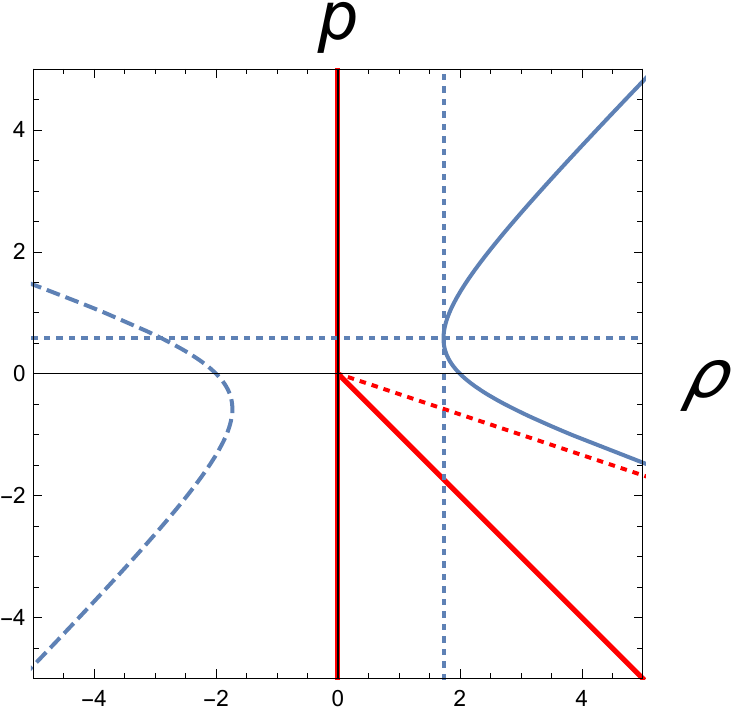}}
  \caption{At the branching point the energy density reaches its minimum. The physical part of the phase curve is in Region Ia so the acceleration is always negative.  }
\label{figure10}
  \end{figure}

For the sake of completeness we discuss also a case where the energy density has a minimum value.
This is most simply obtained by changing a sign in the Lagrangian (\ref{lagneg}):
\begin{equation}
F =   X - \frac 1 X ,  \ \ \ \  \rho= \frac{X^2+3}{2 X},  \ \ \ \ p= \frac{1}{2} \left(X-\frac{1}{X}\right), \ \ \ \ \rho + 3 p = 2X. \label{eos3ads{i}}
\end{equation}
The energy density is always positive and has a minimum at $ X_c= \sqrt 3 $; the  acceleration always negative (see Fig. \ref{figure10}). 
The upper  and lower sheets of the phase curve are  given  by 
\begin{equation}
p= \frac 13 \left(\rho \pm 2 \sqrt{\rho^2-3}\right).
\label{rami2ads}
\end{equation}  
  The cosmic evolution is now ruled by the following equations $(a_0=1)$:
\begin{eqnarray}
a(X)=\frac{\sqrt{X}}{\sqrt[3]{X^2+1}}\label{ax35},  \ \ \ 
  \dot X =\pm\frac{\sqrt{6} X \sqrt{X+\frac{3}{X}} \left(X^2+1\right)}{X^2-3}. \label{eosi35}
\end{eqnarray}

\begin{figure}[h]
\center{\includegraphics[height =5cm]{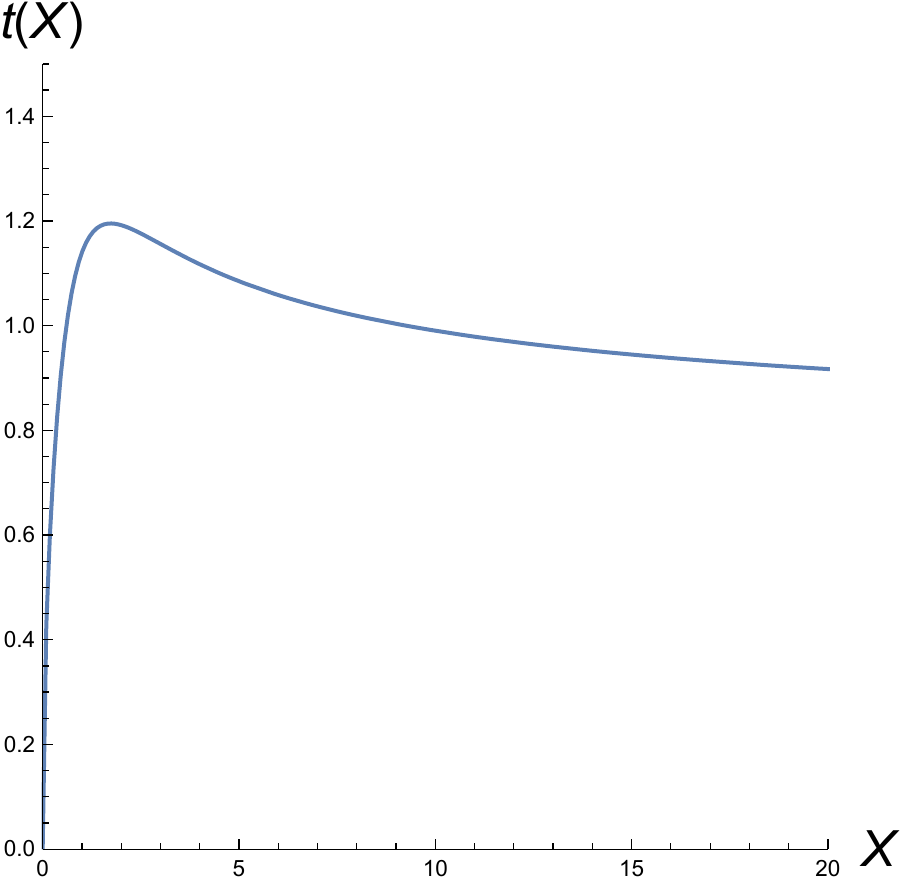}    \ \ \ \ \ \ \  \includegraphics[height=5cm]{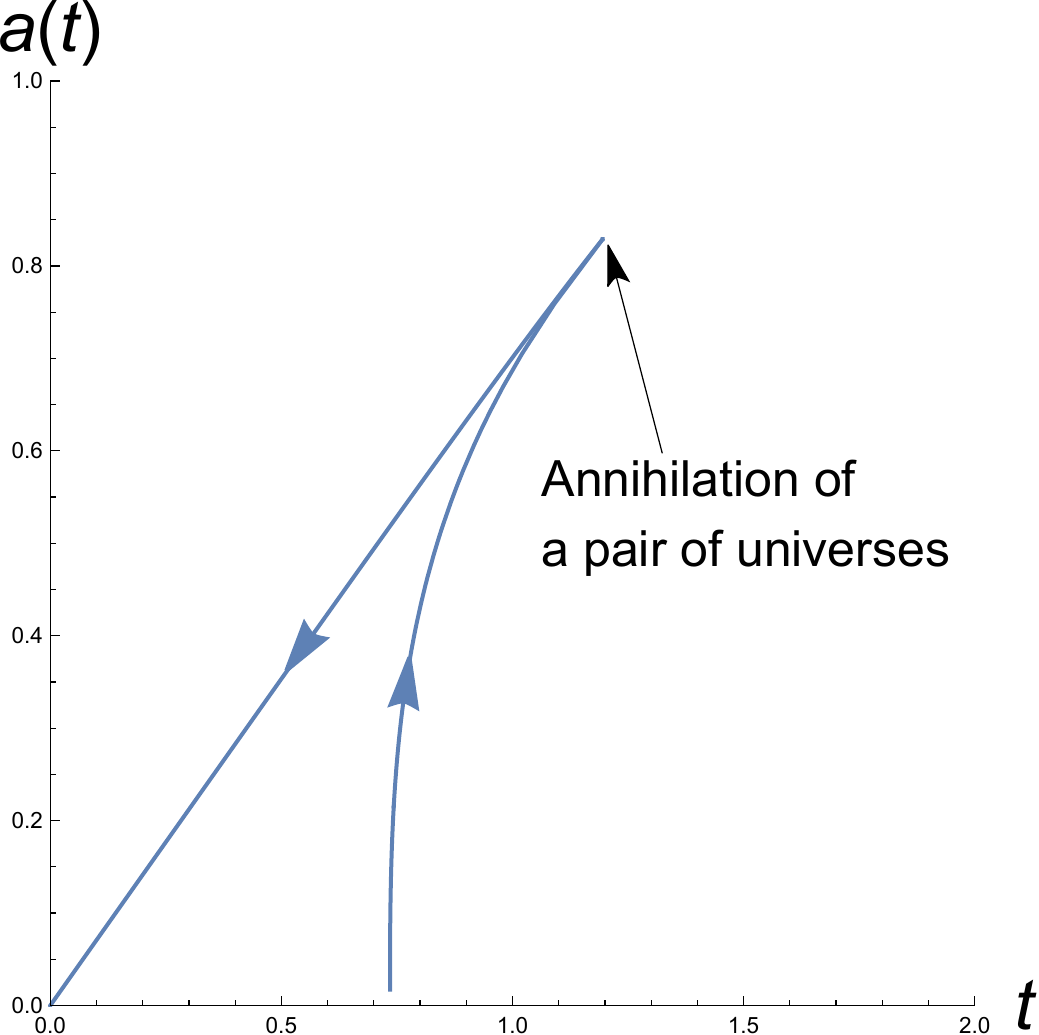}  }
\caption{The energy density has a minimum when of a pair of universes annhilates.}
 \label{fig11}
\end{figure}

By giving initial conditions at $X\sim\infty$ on the stable upper branch, $\dot X$ has to be chosen negative and therefore 
\begin{eqnarray}
\dot a(X)=\frac{\sqrt{X^2+3}}{\sqrt{6} \sqrt[3]{X^2+1}} >0 , \ \ \ \ \ \ddot a(X) =\frac{X^{3/2}}{3 \sqrt[3]{X^2+1}}<0.
\end{eqnarray}
The universe starts from an initial singularity; it expands decelerating and gets to the branching point $X_c=\sqrt 3$ after  a finite interval of cosmic time
\begin{eqnarray}
T= \int_{\sqrt{3}}^\infty\frac{X^2-3}{\sqrt{6 X} \sqrt{X^2+3} \left(X^2+1\right)}dX.
\end{eqnarray}
At $X_c$ the energy density has a minimum so the universe cannot expand anymore. But the velocity $\dot a (X_c)= {1}/{\sqrt[3]{4}}$ is positive so the universe cannot stay at the branching point as well. Again the function $X(t)$ obtained by inverting 
\begin{eqnarray}
t(X)=2 \sqrt{2X}
   F_1\left(\frac{1}{4};-\frac{1}{2},1;\frac{5}{4};-\frac{X^2}{3},-X^2\right)-\sqrt{2X} \,
   _2F_1\left(\frac{1}{4},\frac{1}{2};\frac{5}{4};-\frac{X^2}{3}\right) \end{eqnarray}
is ramified  (see Fig. \ref{fig11}).
In the second branch the direction of time is inverted and the universe starts contracting  to get to the singularity at $X=0$ in a finite interval of cosmic time.

\section{General models}
\label{6}
\begin{figure}[h]\centerline{   \includegraphics[height=7cm]{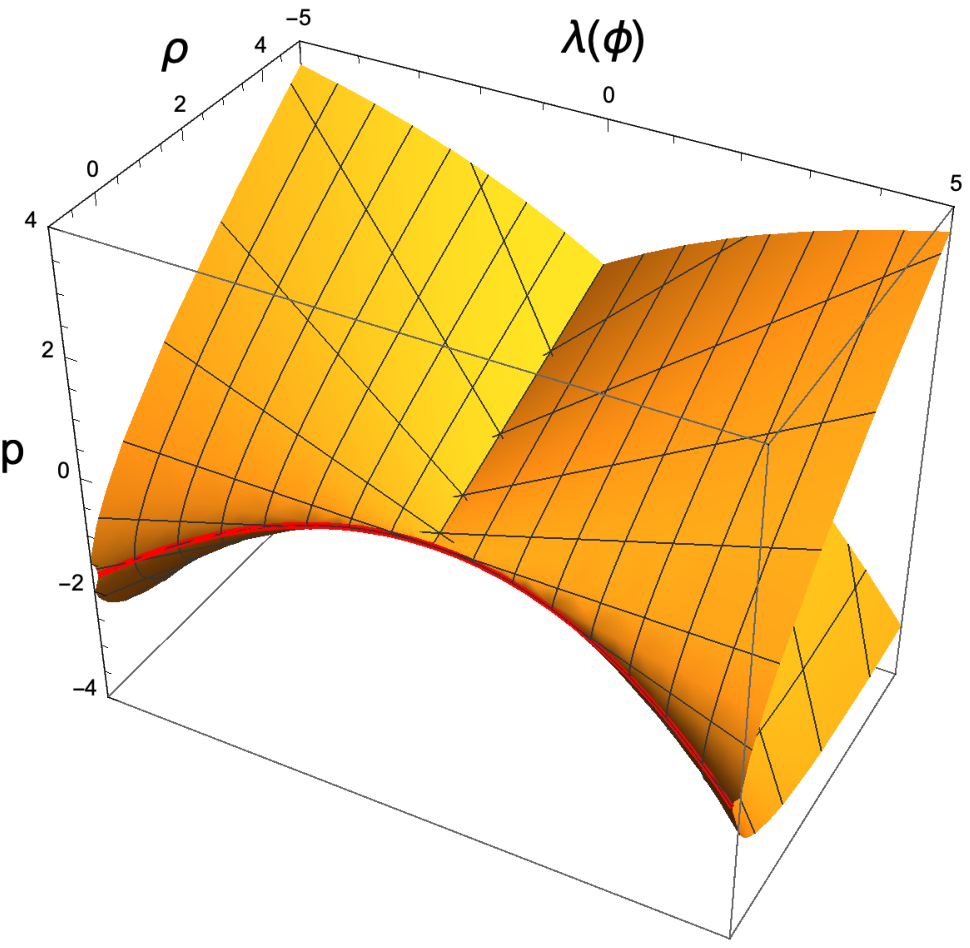}\ \ \ \  \includegraphics[height=6cm]{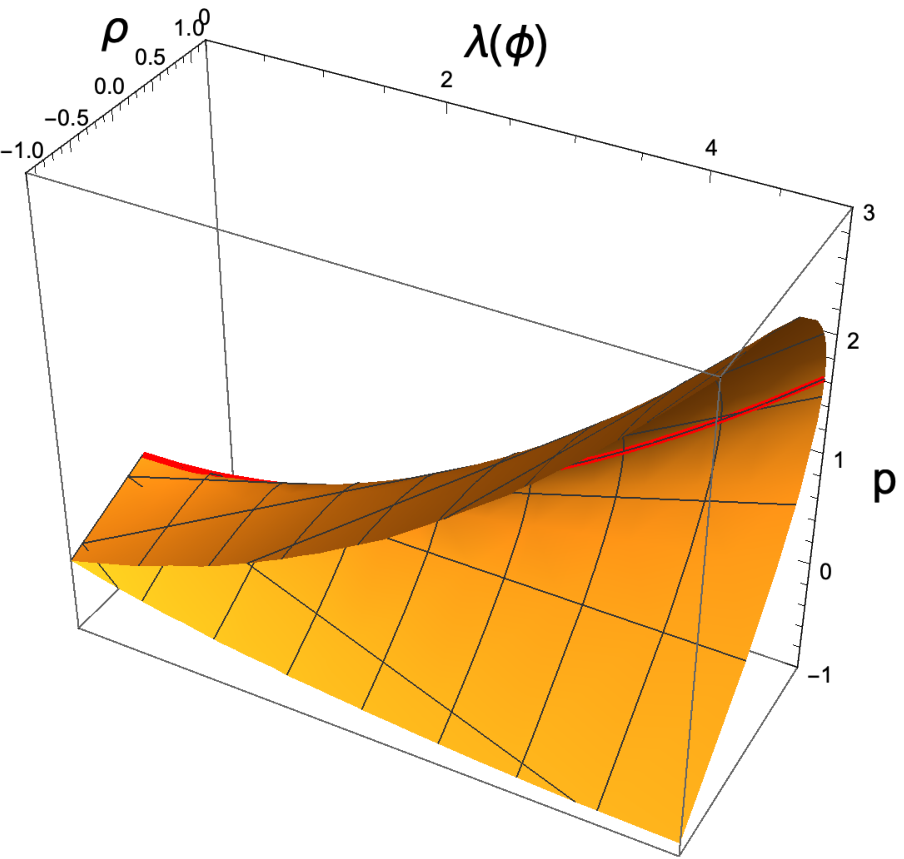}}
  \caption{At the left the equation of state for models  of class I at the right for models of class II. The branching curve is the red curve. At the branching curve the energy density reaches its maxima and the effective speed of sound diverges. }
\label{figeos345}
  \end{figure}

At this point one may ask whether the strange behaviour of a Friedmann-Lema\^itre universe going trough a ramification point of an equation of state  is just a peculiarity of the purely kinetic $k$-essential models that have not enough  degrees of freedom to avoid it.
To show that the features described in Sec. \ref{sec2} are indeed generic  we turn now our attention to the Lagrangian (\ref{theo}) in its full generality. By suitable field redefinitions \cite{damour} we are left with the following three possibilities:
\begin{eqnarray}
{\rm I.} && F_1 = \lambda(\varphi)  X + X^2 =  \lambda(\varphi)  \dph^2 + \dph^4 , \\
{\rm II.}&& F_2 = \lambda(\varphi)  X - X^2 =  \lambda(\varphi)  \dph^2 - \dph^4 , \ \ \ \ \lambda(\varphi)>0.\\
{\rm III.}&& F_3 = X +\mu(\varphi) X^2 =    \dph^2 + \mu(\varphi)  \dph^4  .
\end{eqnarray}
In models  of the first and the third classes the coupling functions $\lambda(\varphi)$ and $\mu(\varphi)$ are allowed to vanish and to change their signs \cite{damour,Vikman}; $\lambda(\varphi)$  is bound to be strictly positive  in models of the second class. 
The first class of models was introduced in the seminal paper \cite{damour} to describe $k$-inflation for various choices of the coupling $\lambda(\varphi)$.  Here the critical curve is outside  the physical region with the only exception of the origin $\lambda = 0, \rho = 0$ (see Fig. \ref{figeos345}). 
\vskip10pt
Let us  focus on the second class to which the model (\ref{eos3}) belongs as a very special simple case. Energy and pressure are given by 
\begin{eqnarray}
&& \rho= \frac{1}{2} \left( \lambda(\varphi)   \dph^2 -3 \dph^4\right), \ \ \ \ p= \frac{1}{2} \left( \lambda(\varphi)  \dph^2 -\dph^4\right).
\label{eos3bis}
\end{eqnarray}
The energy density becomes negative for $\dot \varphi^2> \lambda(\varphi)/3 $. Given $\varphi$, the maximum of the energy density is attained at at $X_c(\varphi) =\dph^2_c= \lambda(\varphi)/6$:
\begin{equation}
\rho_{max}(\varphi)=\lambda(\varphi)^2/24. \label{maxima}
\end{equation} 
The condition $\rho + p\geq0$ is violated for $X>\lambda(\varphi)/2$. 

 The ramified equation of state (\ref{eoszero}) is here written as follows: 
 \begin{equation}
p(\rho,\varphi)= \frac{1}{18} \left(6 \rho+ \lambda(\varphi)^2 \pm   \lambda(\varphi) \sqrt{ \lambda(\varphi)^2-24 \rho }\right).
\label{rami2}
\end{equation}
The maxima of $\rho(\varphi)$ (at given $\varphi$) constitute  the ramification curve for the surface (\ref{rami2}) (the red curve in Fig. (\ref{figeos345})); the upper and lower sheets are respectively given by the upper and lower signs at the rhs of Eq. (\ref{rami2}). At the maxima, the  effective squared speed of sound 
\begin{eqnarray}
 c_s^2 = \frac {\partial p }{\partial \rho }= \frac{1}{3}\mp \frac{2 \lambda(\varphi)}{3 \sqrt{\lambda(\varphi)^2-24 \rho}}  \label{sound2}
\end{eqnarray}
diverges. The effective squared speed of sound has also been given the following expression   \cite{mu} which is identical to (\ref{sound2}):
\begin{eqnarray}
 c_s^2 = \frac {\frac{\partial p }{\partial X}}{\frac{\partial \rho }{\partial X}}= \frac{\lambda(\varphi)-2 \dot\varphi^2}{\lambda(\varphi)-6 \dot\varphi^2}.   \label{sound1}
\end{eqnarray}
$ c_s^2$  negative for  $\frac{ \lambda(\varphi) } 6 <\dot\varphi^2<\frac { \lambda(\varphi) }2$; this contains  the "physical" part  of the upper sheet of the phase surface (\ref{rami2})  that lies in the physical region where both $\rho$ and $\lambda$ are positive (see Fig. (\ref{figeos345})) and of course  diverges at $X=X_c(\varphi)$.

\newpage
The field equations are 
 \begin{eqnarray}  
&& \ddot\varphi + 3 H\frac{\lambda(\varphi) - 2\dph^2}{\lambda(\varphi) - 6\dph^2} \dph +   \frac 1{2(\lambda(\varphi) - 6\dph^2) } \frac{\partial \lambda }{\partial \varphi} \dph ^2 =0.  \label{equ2}
\\ && \cr && \cr
&&H= \frac{\dot a}{a} = 
 \pm \sqrt{\frac \rho  3}=\pm    \sqrt{\frac 1 6\dph^2( \lambda(\varphi) -3  \dph^2), }\label{equ2b}\\
&& \frac{\ddot a}{a} = 
-\frac 1 6 (\rho +3p) =\frac{1}{2} \dph^4 -\frac{1}{3} \lambda (\varphi ) \dph ^2 ,   \label{equ3}
\end{eqnarray}
They might be used to reverse-engineering the coupling $\lambda(\varphi)$ for a given  time evolution of the universe but we will not consider this possibility here. 
In any case, solving them analytically is out of the question even for very simple choices of the coupling $\lambda(\varphi)$.
 However, the exact solution for the case $\lambda=1$, discussed at length  in  the previous section, will guide us  in performing a qualitative analysis.

\subsubsection*{Lower sheet of the phase surface}
First of all we notice that  setting the initial condition $\dph= 0$ at $t=t_i$ gives the trivial constant solution $\varphi = \varphi_i$. This implies that  the value $\dph= 0$ cannot be attained at finite values of the cosmic time but only in the limits  $t \to \pm \infty$ and, unless a discontinuity is traversed, $\dph$ does not change its sign. 
We shall therefore set 
\begin{equation}
\varphi(t_i) = \varphi_i >0,\ \ \ \ \ \dph(t_i)= \epsilon>0 \ \ \  at  \ \ t=t_i>-\infty. 
\end{equation}
Since $\dph(t_i)^2  \sim  \epsilon^2 <\lambda (\varphi_i)/6$ and  because  $\lambda(\varphi)$ is  strictly positive, the above initial conditions locate the system in the lower sheet of the phase surface (\ref{rami2}).   
As in Eq. (\ref{acc}) let us suppose at first that the universe is contracting and choose the minus sign at the rhs of Eq. (\ref{equ2b}) :
 \begin{eqnarray}  
\ddot\varphi =  \sqrt{\frac 3 2( \lambda(\varphi) -3  \dph^2)} \frac{\lambda(\varphi) - 2\dph^2}{\lambda(\varphi) - 6\dph^2} \dph^2 -   \frac 1{2(\lambda(\varphi) - 6\dph^2) } \frac{\partial \lambda }{\partial \varphi} \dph ^2 . \label{equ2m}
\end{eqnarray}
From Eq. (\ref{equ2m}) one sees that $\ddot\varphi(t_i) >0$ provided that 
$ {\partial_\varphi  \lambda} (\varphi_i) \leq \sqrt 6 (\lambda(\varphi_i) )^{3/2} . $
The slightly more restrictive condition 
\begin{equation}
 {\partial_\varphi  \lambda} (\varphi) \leq  \frac 5 {2\sqrt 6}\lambda(\varphi )^{3/2} \simeq 1.02062 \, \lambda (\varphi)^{3/2}\label{condu}
 \end{equation}
guarantees   that $\ddot \varphi(t)>0$ as long as the dynamical system stays in the lower sheet of the phase surface;  this implies that $\dot \varphi(t)$ is  strictly  increasing as  the cosmic time and consequently  also $\varphi(t)$ increases. From 
\begin{equation}
\frac 12 \partial_t ({\lambda(\varphi) - 6\dph^2} )^2=  \dot\varphi {\lambda (\varphi)  \lambda '(\varphi)} - 6 \sqrt{6} \dot\varphi^3 {\sqrt{\lambda (\varphi)-3 \dot\varphi^2} \left(  \lambda (\varphi)-2 \dot\varphi^2   \right)}
\end{equation}
we deduce that, if 
\begin{equation}
 {\partial_\varphi  \lambda} (\varphi) <\frac{2 \, \lambda (\varphi)^{3/2}}{\sqrt{3}},
\end{equation}
the quantity $({\lambda(\varphi) - 6\dph^2} )$ is positive and strictly decreasing on lower sheet of the phase surface (\ref{rami2})  and eventually vanishes at a certain time $t_c$; at that point   $X$ reaches the critical value and $\dot X$ diverges.

Altogether, if the {\em sufficient condition} (\ref{condu}) holds, as the cosmic time  approaches $t_c$  the quantity   $\ddot \varphi(t)$ tends to infinity  and the curve $\dot \varphi (t)$ becomes  vertical exactly as in the purely kinetic case (see Eq. (\ref{uhu}) and Figs. \ref{fighe3} and \ref{fighe4}). At the critical point 
\begin{equation}
\lambda(\varphi(t_c)) = 6\dph(t_c)^2, \ \ \ \ \frac{\dot a(t_c)}{a(t_c)} = -\frac {\dot \varphi(t_c) }{\sqrt 2},\ \ \ \ \frac{\ddot a(t_c)}{a(t_c)} = -\frac {3 \dot \varphi(t_c) ^2}{ 2}.
\end{equation}
\vskip 10pt
\subsubsection*{Upper sheet of the state equation}
\begin{figure}[h]\centerline{ \includegraphics[width=5cm]{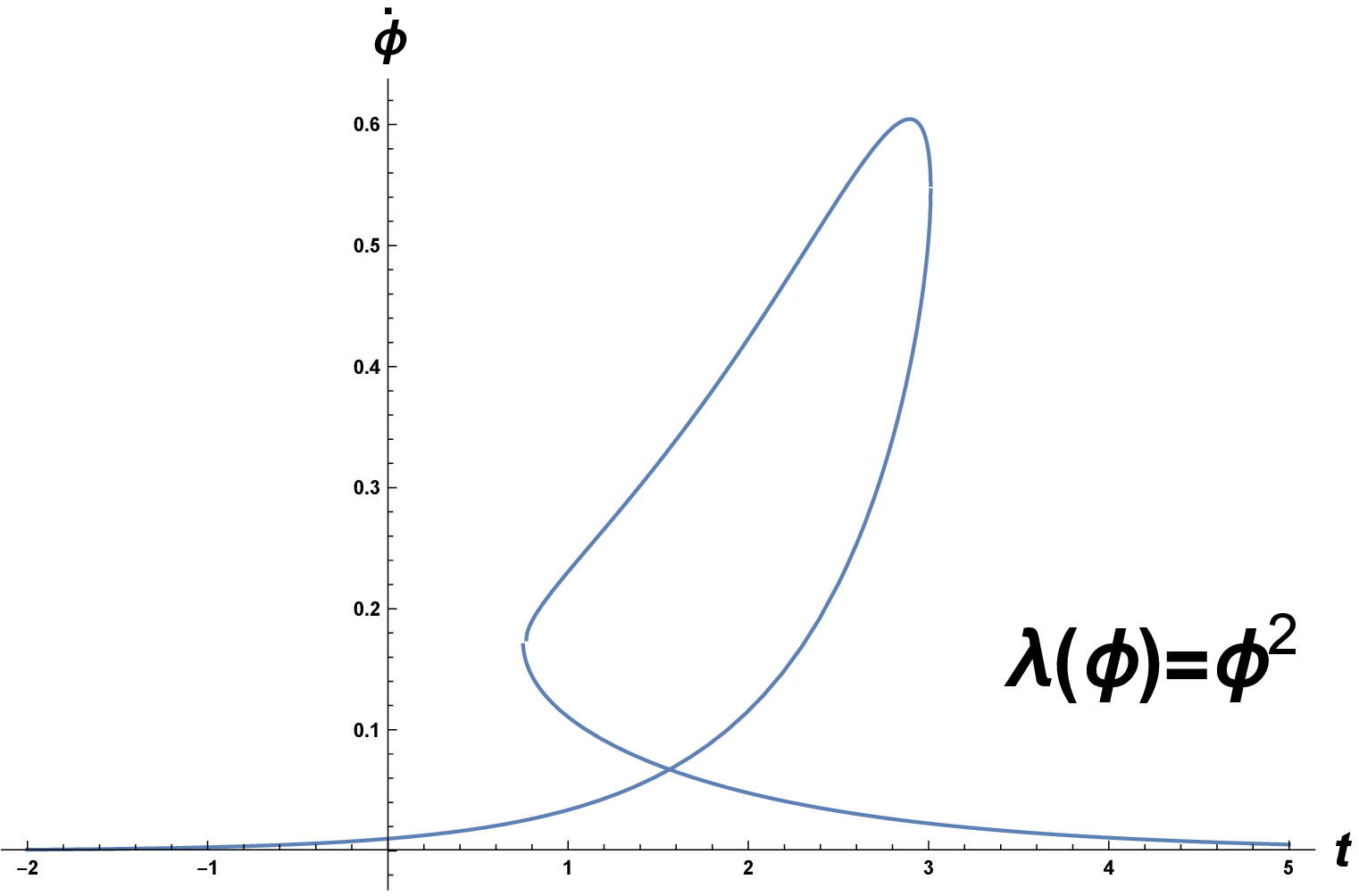} \ \   \includegraphics[width=5cm]{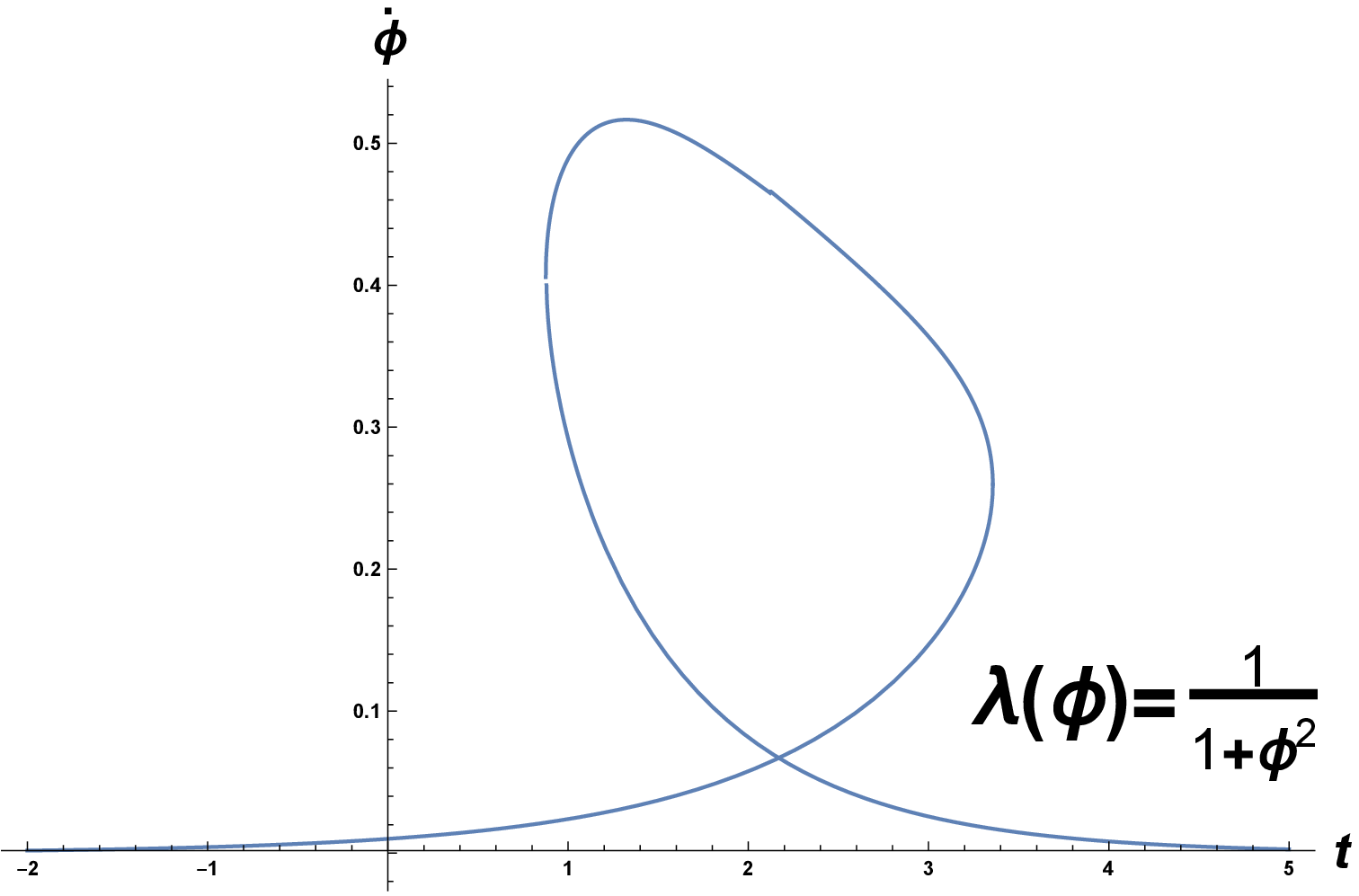} \ \
\includegraphics[width=5cm]{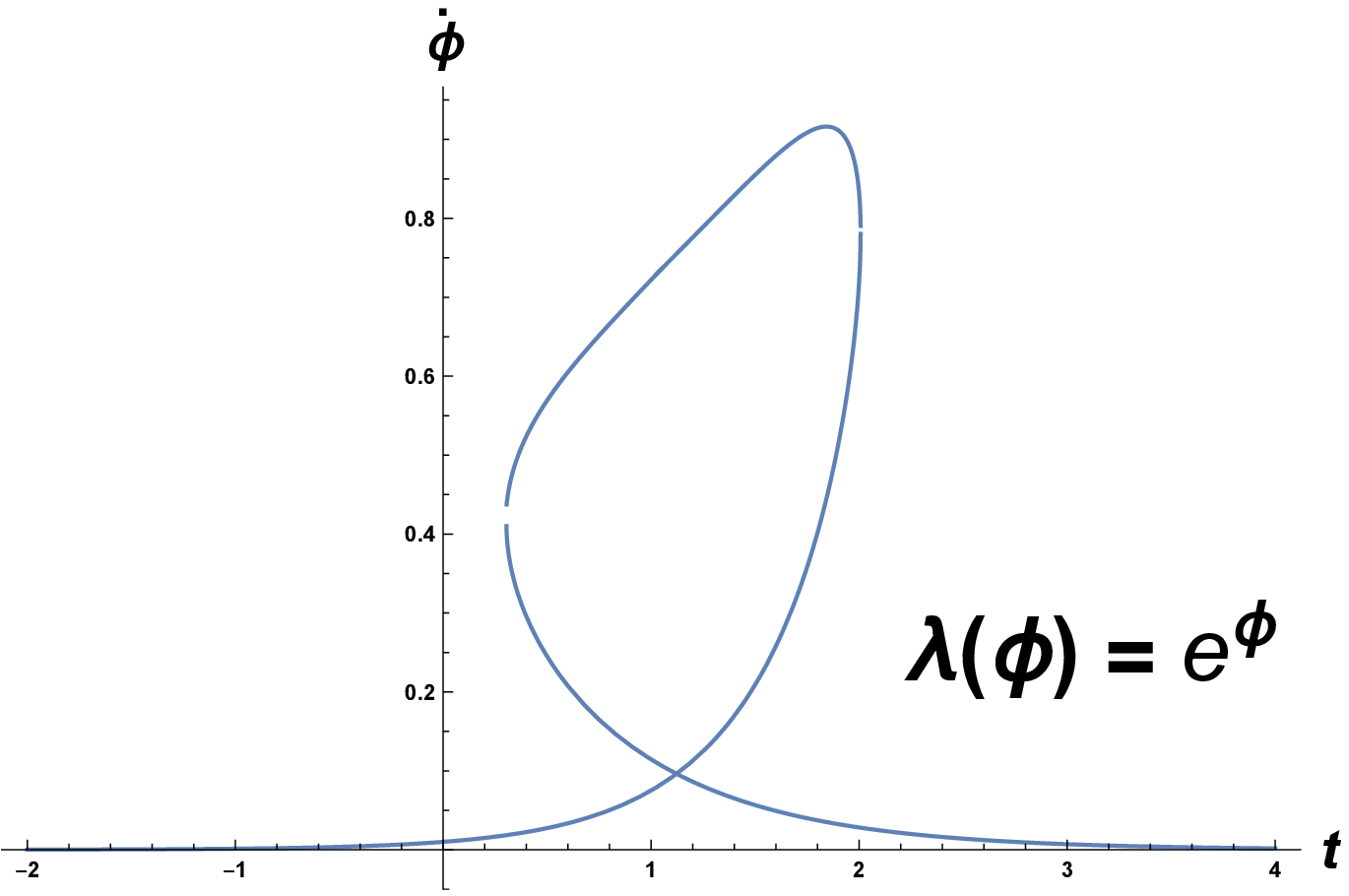}}
  \caption{Generic solutions for three different choices of the coupling $\lambda(\varphi)$. The chosen initial condition are $\varphi(0) = 1, \dot\varphi(0) = 0.01$. The maxima of $\dot \varphi $ are attained in the interior of the upper sheet of the phase surface and therefore the time loops are no more symmetric. The positivity of the energy density is protected by the boundary  $\lambda (\varphi) =  3 \dot\varphi^2$. The dynamics changes very little for quite different choices of potential and is essentially driven by the boundaries of the phase surface and by the critical curve. Of course the conservative interpretation of the time evolution without time loops is also available.}
\label{figure10q}
  \end{figure}

Let us now give initial conditions at $\lambda (\varphi) =  3 \dot\varphi^2$  i.e. at  the boundary of the upper sheet of the phase surface (\ref{rami2}) where $\rho = 0$:
\begin{equation}
\varphi(t_{i}) = \varphi_{i} ,\ \ \ \ \ \dph(t_{i})= \sqrt{\frac{\lambda(\varphi_{i})}{3}}\ \ \  at  \ \ t=t_{i}. 
\end{equation}
Since  $\dot a (t_{i})=0$ and  ${\ddot a (t_{i})} /{a (t_{i})}=-{\lambda{(\varphi}_{i})}/ {54}<0$,  
from   Eq. (\ref{equ2}) it follows  that 
  \begin{eqnarray}  
\ddot\varphi(t) = \left(\mp  \sqrt{\frac 3 2( \lambda(\varphi(t)) -3  \dph(t)^2)} \frac{\lambda(\varphi(t)) - 2\dph(t)^2}{\lambda(\varphi(t)) - 6\dph(t)^2} -   \frac 1{2(\lambda(\varphi(t)) - 6\dph(t)^2) } \frac{\partial \lambda }{\partial \varphi(t)}\right) \dph(t) ^2 , \cr t \  {\lessgtr} \ t_{i}.
\end{eqnarray}
 Note that $\ddot \varphi$  is continuous at $t=t_{i}$:  
  \begin{eqnarray}  \ddot\varphi (t_i) =  \frac 1 6  \frac{\partial \lambda }{\partial \varphi} (\varphi_i).
\end{eqnarray}
When $\lambda(\varphi)$ is a constant, as in the purely kinetic case, $\dot \varphi$ has an extremum on the boundary $\lambda (\varphi) =  3 \dot\varphi^2$  of the phase surface (see Figs. \ref{Fighe2},  \ref{fighe3} and \ref{fighe4}). In general this is not true when $\lambda(\varphi)$ is not a constant. 
This behaviour is exemplified in the plots of Fig. \ref{figure10q} where $\dot \varphi$ is numerically solved and plotted for the couplings $\lambda(\varphi)=\varphi^2$ , $\lambda(\varphi)=1/(1+\varphi^2)$ and $\lambda(\varphi)=\exp\varphi)$. It is seen that the dynamics changes very little and is eesntiaky driven by the boundaries of the phase surface and by the critical curve.

However, since  
\begin{equation}
\partial_t \rho(t) = \mp \sqrt{\frac{3}{2}} \dot \varphi (t)^3 \sqrt{\lambda (\varphi (t))-3 \dot\varphi
   (t)^2} \left(\lambda (\varphi (t))-2\dot  \varphi (t)^2\right),  \ \ \ \ \ t \  {\lessgtr} \ t_{i}, \label{pp00}
   \end{equation}
it follows that 
\begin{equation}
\rho(t_i)=0, \ \ \ \partial_t \rho(t_i)=0 ,\ \ \   \partial^2_t \rho(t_i)=\frac{1}{54} \lambda (\phi (t))^4;
   \end{equation} this means that the energy density cannot become negative whatever is the choice of the coupling $\lambda(\varphi)$ and  the boundary of the upper sheet where the density vanishes  cannot be trespassed;   the crucial fact is that Eqs. (\ref{pp00}) do not depend on ${\partial \lambda }/{\partial \varphi} $. 

\section{Summary}

We have examined the flat FLRW cosmological models based on the quartic $k$-essence  Lagrangian (\ref{theo}) and  
discussed  features that have stayed uncovered to date, as they  cannot be qualitatively guessed by inspecting the equation of state even in the simpler purely kinetic case.  They are related to the existence of boundaries and branching curves  in the phase surfaces of the models, a (mathematical) feature that seems to be completely overlooked in the literature. 

The  dynamical properties of  a cosmological model traversing a branching point are quite strange; there is also relation to the arrow of time that we found interesting and new;  it is described in Sections  \ref{sec2} and \ref{6} .
Actually these features are not specific to  $k$-essential models but are expected to always show up when the equation of state is a multivalued function of the thermodynamical parameters.
It remains to understand whether our universe has ever met a branching point and, if yes, what would be the observable consequences of this fact. This would require  to understand how branching points may be reconciled with the standard wisdom of linear cosmological perturbations, a question that we leave for future work.

\section*{Acknowledgments}
U. M. is indebted to Thibault Damour for many illuminating discussions and suggestions and to the IHES for its generous hospitality during the writing of this paper. \\
M.N. thanks  CNPq and FAPERJ for financial support. 

\end{document}